\documentclass[10pt]{iopart}

\usepackage{mathrsfs}
\usepackage{iopams}
\usepackage{cite}
\usepackage{epsfig}
\usepackage{graphicx}
\usepackage{color}

\def\Th{\Theta}
\def\sig{\sigma}
\def\om{\omega}
\def\udot{\dot{u}}
\def\3nab{\tilde{\nabla}}
\def\sq{\square}

\def\lgl{\langle}
\def\rgl{\rangle}

\def\hsp5{\hspace{5mm}}
\newcommand{\sfrac}[2]{{\textstyle{#1\over#2}}}
\def\case#1/#2{\textstyle\frac{#1}{#2}}

\def\be {\begin{equation}}
\def\ee {\end{equation}}
\def\bea {\begin{eqnarray}}
\def\eea {\end{eqnarray}}

\def\bi {\bibitem}
\def\case#1/#2{\textstyle\frac{#1}{#2} }
\def\rf#1{(\ref{#1})}
\def\cqg{{\em Class. Quantum Grav.\/} }
\def\grg{{\em Gen. Rel. Grav.\/} }
\def\prd{{\em Phys. Rev.\/} {\bf D}}
\def\prl{{\em Phys. Rev. Lett.\/} }
\def\apj{{\em Astrophys. J.\/} }
\def\jmp{{\em J. Math. Phys.\/} }

\begin{document}

\title{Shear dynamics in Bianchi I cosmologies with $R^n$-gravity}

\author[J.A. Leach {\it et al.}]{Jannie A Leach \dag, Sante Carloni \dag \\
and \ Peter K S Dunsby  \dag \ddag } \vspace{3mm}

\address{\dag \ Department of Mathematics and Applied Mathematics, University of Cape Town, Rondebosch,
7701, South Africa}

\address{\ddag \  South African Astronomical Observatory, Observatory, Cape Town, South Africa}

\date{\today}

\eads{leachj@maths.uct.ac.za, scarloni@maths.uct.ac.za and
pksd@maths.uct.ac.za}

\begin{abstract}
We give the equations governing the shear evolution in Bianchi
spacetimes for general $f(R)$-theories of gravity. We consider the
case of $R^n$-gravity and perform a detailed analysis of the
dynamics in Bianchi I cosmologies which exhibit {\it local
rotational symmetry}. We find exact solutions and study their
behaviour and stability in terms of the values of the parameter $n$.
In particular, we found a set of cosmic histories in which the
universe is initially isotropic, then develops shear anisotropies
which approaches a constant value.
\end{abstract}


\section{Introduction}

In the last few years there has been renewed interest in theories of
gravity where the gravitational Lagrangian is a non-linear function
of the scalar curvature. These $f(R)$-{\it theories of gravity} can
take on a number of forms, the majority of the functions considered
being of the type $R+\epsilon R^m$. Theories with $m=-1$ have been
proposed as possible alternatives to sources of dark energy to
explain the observed cosmic acceleration \cite{Carroll04,Nojiri03}.
Solar system experiments do however constrain these type of theories
for any corrections higher than $R^2$ (quadratic gravity)
\cite{Olmo05}. In these theories corrections to the characteristic
length scale of General Relativity (GR) are introduced through the
addition of a new length scale which is determined by the constant
$\epsilon$.

There are however forms of $f(R)$ which do not alter the
characteristic length scale, for example $R^n$, in which GR is
recovered when $n=1$. These $R^n$-gravity theories have many
attractive features, such as simple exact solutions which allows for
comparison with observations \cite{Capozzi02,Capozzi03}. There are
however some caveats, in particular the stability and global
behaviour of the underlying cosmological model is not well
understood. The dynamical systems approach \cite{Dynamical} can
address some of these problems, since it provides one with exact
solutions through the determination of fixed points and a
(qualitative) description of the global dynamics of the system.
Carloni {\it et al} \cite{Carloni05} have recently used this method
to study the dynamics of $R^n$-theories in
Friedmann-Lema\^{i}tre-Robertson-Walker (FLRW) universes. Clifton
and Barrow \cite{Clifton05} used the dynamical systems approach to
determine the extent to which exact solutions can be considered as
attractors of spatially flat universes at late times. They compared
the predictions of these results with a range of observations and
were able to show that the parameter $n$ in FLRW may only deviate
from GR by a very small amount ($n-1\sim 10^{-19}$).

The dynamics of anisotropic models with $f(R)$-gravity have not been
studied as intensively as their FLRW counterparts and it is
therefore not known how the  behaviour of the shear is modified in
these theories of gravity. Bianchi spacetimes with isotropic
3-surfaces have been investigated for the quadratic theory
\cite{Maartens94} and it was found that in Bianchi I cosmologies the
universe isotropises slower than in the Einstein case. The equations
governing the evolution of shear in Bianchi spacetimes for general
$f(R)$-theories can be found from the trace-free Gauss-Codazzi
equations (see for example \cite{carge73,Cargese}). However, it is
not easy to solve these equations since the shear depends
non-linearly on the Ricci scalar. Consequently the dynamical systems
approach provides us with the best means of understanding the
dynamics of these models.

In GR the vacuum Kasner solutions \cite{Kasner} and their fluid
filled counterparts, the Type I Bianchi models, proved useful as a
starting point for the investigation of the structure of anisotropic
models. Barrow and Clifton \cite{Barrow06,Clifton06a} have recently
shown that it is also possible to find solutions of the Kasner type
for $R^n$-gravity models. In \cite{Barrow06} they showed that exact
Kasner-like solutions do exist in the range of parameter $n$ for
$1/2<n<5/4$ but with different Kasner-index relations to the ones in
GR. In this paper we extend the dynamical systems analysis of
$R^n$--gravity \cite{Carloni05} to Bianchi I cosmological models
that exhibit {\it local rotational symmetry} (LRS)
\cite{Ellis67,Stewart68,vElst96}. LRS spacetimes geometries are
subgroups within anisotropic spacetimes in which isotropies can
occur around a point within the spacetime in 1- or 3--dimensions.
Thus there exists a unique preferred spatial direction at each point
which constitutes a local axis of symmetry. All observations are
identical under rotation about the axis and are the same in all
spatial directions perpendicular to that direction
\cite{Ellis67,Stewart68}.

Our analysis of the vacuum case revealed some interesting features
regarding the evolution of Bianchi I cosmologies in HOTG. The phase
space contains one isotropic fixed point and a line of fixed points
with shear. The isotropic fixed point is an attractor (stable node)
for values of the parameter $n$ in the ranges $n<1/2$, $1/2<n<1$ and
$n>5/4$. In the range $1<n<5/4$ this point is a repeller (unstable
node) and therefore may be seen as a past attractor.  The existence
of an isotropic past attractor implies, as in the case of braneworld
models \cite{Coley02a,Coley02b,Dunsby04,Goheer04}, that we do not
require special initial conditions for inflation to start since the
cosmological singularity is FLRW. Another feature of these models is
that the shear evolution is independent of the value of $n$ and is
the same for all fixed points on the line.

The outline of this paper is as follows: In section 2 we give the
basic equations for the kinematical and dynamical variables and we
also state the field equations for general $f(R)$-theories of
gravity. In section 3 we convert these equations into an autonomous
set of equations for the case of $R^n$. We then proceed in section 4
and 5 to analyse this system for models with vacuum and matter
respectively.

The following conventions will be used in this paper: the metric
signature is $(-+++)$; Latin indices run from 0 to 3; $\nabla$
represents the usual covariant derivatives which may be split
($1+3$--covariantly) with the spatial covariant derivative being
denoted by $D$ and the time derivative by a dot; units are used in
which $c=8\pi G=1$.

\section{Kinematics and dynamics of $f(R)$-gravity}

\subsection{Kinematical and dynamical quantities}

We first state the relevant equations governing relativistic fluid
dynamics (see for e.g. \cite{carge73,Cargese}). For any given fluid
4-velocity vector field $u^{a}$, the projection tensor
$h_{ab}=g_{ab}+u_{a}u_{b}$ projects into the instantaneous
rest-space of a comoving observer. The first covariant derivative
can be decomposed as
\begin{equation}
 \nabla_{a}u_{b} = -\,u_a\,\udot_b +
D_{a}u_{b} = -\,u_a\,\udot_b + {\sfrac{1}{3}}\,\Th\,h_{ab} +
\sigma_{ab} + \om_{ab} \ , \label{eq:kin}
\end{equation}
where $\sig_{ab}$ is the symmetric shear tensor ($\sig_{ab} =
\sig_{(ab)}$, $\sig_{ab}\,u^b = 0$, $\sig^a{}_{a}= 0$), $\om_{ab}$
is the vorticity tensor ($\om_{ab} = \om_{[ab]}$, $\om_{ab}\,u^b =
0$) and $\udot_a$ is the acceleration vector
($\udot_a=u^b\nabla_{b}u_{a}$). $\Theta$ is the volume expansion
($\Theta = \nabla_au^a$) which defines a length scale $a$ along the
flow lines via the standard relation $\Theta =\sfrac{3\dot{a}}{a}$.

The matter energy-momentum tensor $T^M_{ab}$ can be decomposed
relative to $u^a$ in the form
\begin{equation}
T^M_{ab} = \mu\,u_a\,u_b + q_a\,u_b + u_a\,q_b + p\,h_{ab} +
\pi_{ab} \label{eq:stress}  \ , \nonumber
\end{equation}
where $\mu$ is the relativistic energy density, $p$ the isotropic
pressure, $q^{a}$ the energy flux ($q_a\,u^a = 0$) and $\pi_{ab}$
the trace-free anisotropic pressure ($\pi^a{}_a = 0$, $\pi_{ab} =
\pi_{(ab)}$, $\pi_{ab}\,u^b = 0 $), all relative to $u^a$.

The conservation equations $\nabla^bT^M_{ab}=0$, can be split with
respect to $u^a$ and $h_{ab}$:
\begin{eqnarray}
\dot{\mu} + D_{a}q^{a} +\Th(\mu+p) + 2(\udot_{a}q^{a}) +
(\sig^{a}\!_{b}\pi^{b}\!_{a})=0, \label{eq:cons1}\\
\hspace{-15mm} \dot{q}^{\lgl a\rgl} + D^{a}p + D_{b}\pi^{ab}+
\sfrac{4}{3}\Th\,q^{a} + \sig^{a}\!_{b}\,q^{b} + (\mu+p)\,\udot^{a}
- \udot_{b}\,\pi^{ab} + \om^{ab}q_{b}=0. \label{eq:cons2}
\end{eqnarray}

\subsection{$f(R)$-gravity Field Equations}

In the case where the gravitational Lagrangian is a non-linear
function of the scalar curvature $f(R)$, the action reads
\begin{equation}
{\cal A}=\int dx^4 \sqrt{- g} f(R)+\int {\cal L}_M dx^4,
\label{action:f(R)}
\end{equation}
where ${\cal L}_M$ is the Lagrangian of the matter fields. The
fourth order field equations can be obtained by varying
\rf{action:f(R)}:
\begin{equation}
T^M_{ab}=f' R_{ab}-\sfrac{1}{2}f g_{ab}+
S_{cd}\left(g^{cd}g_{ab}-g^c_{\; a} g^d_{\; b}\right),
\label{field:f(R)T}
\end{equation}
where primes denote derivatives with respect to $R$ and
$S_{ab}=\nabla_a \nabla_b f'(R)$. $S_{ab}$ can be decomposed as
\begin{eqnarray}\label{Sab:f(R)}
\hspace{-20mm}S_{ab} = f'' D_a D_b R + f''' D_a R\; D_b R -
f'''\dot{R}\left(D_b R\;
u_a+D_a R\; u_b \right)+f''' \dot{R}^2\;u_a u_b  \nonumber \\
\hspace{-15mm}- f''\left[D_a\dot{R}\; u_b+u^c\nabla_c(D_b
R)\;u_a\right]  +f''\left[\ddot{R}\;u_a u_b-\dot{R}(D_a
u_b-u_a\dot{u}_b)\right],
\end{eqnarray}
and so the D'Lambertian can be given by
\begin{eqnarray}\label{Sq_S:f(R)}
\hspace{-8mm} S=\sq f'(R) =g^{ab}\nabla_a \nabla_b f'(R) \nonumber\\
\hspace{-5mm} =f'' D^cD_cR+f''' D^c R\; D_c R + f'' \dot{u}^c D_c R
-f''(\ddot{R}+\dot{R}\Th)-f'''\dot{R}^2.
\end{eqnarray}
The field equation (\ref{field:f(R)T}) can be rewritten in the
standard form
\begin{equation}
R_{ab}-\sfrac{1}{2}g_{ab}R=T^{TOT}_{ab}, \label{field:f(R)}
\end{equation}
(when $f'(R)\neq 0$) where the effective stress energy momentum
tensor $T^{TOT}_{ab}$ is given by
\begin{equation}
T^{TOT}_{ab}=f'^{-1}\left[T^M_{ab}+\sfrac{1}{2}g_{ab}\left(f-f'R
\right)+ S_{cd}\left(g^c_{\; a} g^d_{\;
b}-g^{cd}g_{ab}\right)\right]. \label{field:T_tot)}
\end{equation}
As pointed out in \cite{Carloni05}, no matter how complicated the
effective stress energy momentum tensor for the HOTG system is, it
is always divergence free if $\nabla^bT^M_{ab}=0$. The total
conservation equations therefore have the same form as those for
standard matter and can thus be represented by \rf{eq:cons1} and
\rf{eq:cons2}.

The higher order field equations may then be split (see
\cite{carge73,Maartens94,Rippl96}) to give the following
contributions:
\begin{eqnarray}
\hspace{10mm} R&=& f'^{\;-1}\left[3p-\mu+2 f -3\, S\right], \label{f(R):R} \\
R_{ab}u^a u^b &=& f'^{\;-1}\left[\mu-\sfrac{1}{2}f+h^{ab}S_{ab}\right], \label{f(R):Ruu}\\
R_{ab}u^a h^b_c &=& f'^{\;-1}\left[-q_c+S_{ab}u^a h^b_c\right], \label{f(R):Ruh}\\
R_{ab}h^a_c h^b_d &=& f'^{\;-1}\left[\pi_{cd}-\left( p+
\sfrac{1}{2}f+S \right) h_{cd}+ S_{ab}h^a_c h^b_d\right].
\label{f(R):Rhh}
\end{eqnarray}

The propagation and constraint equations for $f(R)$-theories are
given by Ripple {\it et al.} \cite{Rippl96}. In what follows we will
use the {\it Raychaudhuri equation}
\begin{equation}\label{Ray:f(R)}
\hspace{-16mm} \dot{\Th} - D_{a}\udot^{a} + \sfrac{1}{3}\,\Th^{2} -
(\udot_{a}\udot^{a}) + 2\,\sig^{2} - 2\,\om^{2}
+f'^{\;-1}\left[\mu-\sfrac{1}{2}f+h^{ab}S_{ab}\right]=0,
\end{equation}
and the {\it trace free Gauss-Codazzi equation}, which holds for an
irrotational matter fluid flow ($\om_{ab}=0$):
\begin{eqnarray}\label{3GCE:f(R)}
^3R_{ab}- \sfrac{1}{3}({}^3R) h_{ab}&=& -\dot{\sig}_{\lgl
ab\rgl}-\Th\,\sig_{ab} + D_{\lgl a}\udot_{b\rgl}+
\udot_{\lgl a}\, \udot_{b\rgl}+ f'^{\;-1}\pi_{ab} \nonumber \\
&& +f'^{\;-1}\left[h^c_a h^d_b-\sfrac{1}{3}h_{ab}
h^{cd}\right]S_{cd},
\end{eqnarray}
where the 3-Ricci scalar is given by
\begin{equation}\label{3R:f(R)}
^3R = 2\sigma^2-\sfrac{2}{3}\Th^2+f'^{\;-1}\left[\mu+3p+f- S
+2h^{cd}S_{cd}\right].
\end{equation}

\section{Shear dynamics in Bianchi I cosmologies}

\subsection{$f(R)$-gravity}

We consider a Bianchi spacetimes whose homogeneous hypersurfaces
have isotropic 3-curvature $^3R_{ab}=\sfrac{1}{3}({}^3R) h_{ab}$.
These spacetimes include the Bianchi models which, via the
dissipation of the shear anisotropy $\sigma_{ab}$, can reach a FLRW
limit. Spatial homogeneity implies that the spatial gradients will
vanish and that $\udot_a=0=\omega$. Thus the trace free
Gauss-Codazzi equation (\ref{3GCE:f(R)}) becomes
\begin{equation}\label{3GCE:f(R)_Bianchi}
\dot{\sig}_{\lgl ab\rgl}+\Th\,\sig_{ab}=
f'^{-1}\left[\pi_{ab}+\left(h^c_a h^d_b-\sfrac{1}{3}h_{ab}
h^{cd}\right)S_{cd}\right],
\end{equation}
and $S_{ab}$ can be split as follows
\begin{eqnarray}
S_{ab} &=& f''\left(\ddot{R}u_a u_b - \dot{R}\nabla_bu_a\right)+f'''\dot{R}^2u_a u_b, \\
S_{ab}u^a u^b &=& f''\ddot{R}+f'''\dot{R}^2, \\
S_{ab}h^{ab} &=& -f''\dot{R}\Th, \\
S &=&-f''(\ddot{R}+\dot{R}\Th)-f'''\dot{R}^2 \label{f(R):sqS}.
\end{eqnarray}
Substituting these components into the Gauss-Codazzi equation
\rf{3GCE:f(R)_Bianchi} gives
\begin{equation}\label{sigdot:f_Bch1}
\dot{\sig}_{\lgl ab\rgl}+\Th\,\sig_{ab}= f'^{\;-1}\left[\pi_{ab} -
f''\dot{R}\sigma_{\lgl ab\rgl}\right].
\end{equation}
In the case of a perfect matter fluid, $\pi_{ab}=0$, so that the
equation above becomes
\begin{equation}
\dot{\sig}_{\lgl ab\rgl}+\Th\,\sig_{ab}=a^{-3}
\frac{d}{d\tau}\left(a^3\,\sig_{ab}\right)=-\frac{f''\dot{R}}{f'}\sigma_{ab}.
\end{equation}
On integration this yields
\begin{equation}\label{sig_ab:f_Bch1}
\sigma_{ab}=f'^{\;-1}\Psi_{ab}a^{-3},\ \ \ \ \ \dot{\Psi}_{ab}=0,
\end{equation}
which in turn implies
\begin{equation}\label{sig^2:f_Bch1}
\sigma^2=f'^{\;-2}\Psi^2 a^{-6},\ \ \ \ \ \dot{\Psi}^2=0.
\end{equation}
In the case of $f(R)=R$, equation \rf{sig^2:f_Bch1} gives the
standard GR solution (see \cite{Cargese} and references there in)
whose behaviour can be summarised as follows:
\begin{equation*}
\sigma^2 \rightarrow \infty \ \ {\rm as}\ \ a \rightarrow 0,\ \ \ \
\sigma^2 \rightarrow 0 \ \ {\rm as}\ \ a \rightarrow \infty.
\end{equation*}
This behaviour is modified in $f(R)$-theories of gravity (see
\cite{Berkin90,Maartens94}), because $R$ (and therefore $f'$) is a
function of $\sigma^2$ (see \rf{R:gen_B1} below) and therefore
\rf{sig^2:f_Bch1} is implicit. In particular, the dissipation of the
shear in Bianchi I spacetimes is slower in quadratic gravity than in
GR \cite{Maartens94}. However, this result was obtained by solving
the evolution equations under the assumption that the scale factor
also has a power-law evolution. Although this is desirable it may
not necessarily be true since no analytical cosmological solution
could be obtained in \cite{Maartens94}. A more general approach to
this problem is to make use of the theory of dynamical systems (see
\cite{Dynamical} and references therein). In the following we will
apply this technique to $R^n$ gravity in order to investigate
further the behaviour of the shear in this framework.

\subsection{$R^n$-- gravity}

We begin by specialising all the evolution equations above to the
case of $f(R)=R^n$. The Raychaudhuri equation \rf{Ray:f(R)} is now
\begin{equation}
\dot{\Th} + \sfrac{1}{3}\,\Th^{2} + 2\sigma^2 -\frac{1}{2n}R -
(n-1)\frac{\dot{R}}{R}\Th+\frac{\mu}{nR^{n-1}} =0,
\label{Ray:R^n_B1}
\end{equation}
and the trace free Gauss-Codazzi equation \rf{sigdot:f_Bch1} for LRS
spacetimes is given by
\begin{equation}\label{sigdot:Rn_LRS_B1}
\dot{\sig}= -\left(\Th+ (n-1)\frac{\dot{R}}{R}\right)\sigma.
\end{equation}
The Friedmann equation can be found from \rf{3R:f(R)}
\begin{equation}\label{3R:Rn_B1}
\sfrac{1}{3}\,\Th^{2}-\sigma^2+(n-1)\frac{\dot{R}}{R}\Th-\frac{(n-1)}{2n}R-\frac{\mu}{nR^{n-1}}=0.
\end{equation}
In general, the substitution of the Friedmann equation \rf{3R:Rn_B1}
into the Raychaudhuri equation \rf{Ray:R^n_B1} yields
\begin{equation}\label{R:gen_B1}
R=2\dot{\Th}+\sfrac{4}{3}\Th^2+2\sigma^2.
\end{equation}
Note that, in this relation the energy density does not appear
explicitly, but is however still contained implicitly in the
variables on the right hand side.

In this paper we will assume standard matter behaves like a perfect
fluid with barotropic pressure $p=w\mu$. The conservation equation
\rf{eq:cons1} in this case is
\begin{equation}\label{cons:perfect}
\dot{\mu}=-(1+w)\mu\Th.
\end{equation}

In order to convert the equations above into a system of autonomous
first order differential equations, we define the following set of
expansion normalised variables \footnote{It is important to note
that this choice of variables will exclude GR, i.e the case of
$n=1$. See \cite{Dynamical} for the dynamical systems analysis of
the corresponding cosmologies in GR.};
\begin{eqnarray}\label{DS:var}
\Sigma =\frac{3\sigma^2}{\Th^2}\; , \ \ \ \ \ \
&&x = \frac{3\dot{R}}{R\Th}(n-1)\; , \\
y = \frac{3R}{2n\Th^2}(n-1)\; , \ \ \ \ \ \ && z =
\frac{3\mu}{nR^{n-1}\Th^2}\; , \nonumber
\end{eqnarray}
whose equations are
\begin{eqnarray}\label{DS:eqn_mat}
\hspace{-10 mm}
\Sigma'=2\left(-2+2\Sigma-\frac{y}{n-1}-2x+z\right)\Sigma, \nonumber
\\
\hspace{-10 mm} x' = y(2+x)-\frac{y}{n-1}(2+nx)-2x-2x^2+xz+(1-3w)z+2x\Sigma,  \\
\hspace{-10 mm}y'=\frac{y}{n-1}\left[(3-2n)x-2y+2(n-1)z+4(n-1)\Sigma+2(n-1)\right], \nonumber \\
\hspace{-10 mm} z' = z\left[
2z-(1+3w)-3x-\frac{2y}{n-1}+4\Sigma\right] , \nonumber
\end{eqnarray}
where primes denote derivatives with respect to a new time variable
$\tau=\ln a$ and the dynamical variables are constrained by
\begin{equation}\label{DS:constraint_mat}
1-\Sigma+x-y-z=0.
\end{equation}

\section{Dynamics of the vacuum case}

We first consider the vacuum case ($\mu=0$). In this case the set of
dynamical equations \rf{DS:eqn_mat} are given by
\begin{eqnarray}\label{DS:eqn1}
\Sigma'&=& 2 \left(-2+2\Sigma-\frac{y}{n-1}-2x\right)\Sigma,
\nonumber \\
x' &=& y(2+x)-\frac{y}{n-1}(2+nx)-2x-2x^2+2x\Sigma, \\
y'&=&\frac{y}{n-1}\left[(3-2n)x-2y+4(n-1)\Sigma+2(n-1)\right],
\nonumber
\end{eqnarray}
together with the constraint equation
\begin{equation}\label{DS:constraint}
1-\Sigma+x-y=0.
\end{equation}

\subsection{Fixed points and solutions}

The two most useful variables are $\Sigma$ and $y$ since they
respectively represent a measure of the expansion normalised shear
and the expansion normalised Ricci curvature and hence allow us to
investigate how the shear is modified by the curvature. We can
therefore simplify the system \rf{DS:eqn1}, by making use of the
constraint \rf{DS:constraint}, which allow us to write the equation
for $x$ as a combination of the two variables $\Sigma$ and $y$:
\begin{eqnarray}\label{DS:eqn2}
\Sigma' &=& -2\left(\frac{2n-1}{n-1}\right)\; y\; \Sigma, \nonumber \\
y'&=&\frac{y}{n-1}\left[(2n-1)\Sigma-(2n-1)y+(4n-5)\right],
\end{eqnarray}
which together with the constraint \rf{DS:constraint} represents our
new system. Setting $\Sigma'=0$ and $y'=0$ we obtain one isotropic
fixed point $\mathcal{A}:\ \left(0,\sfrac{4n-5}{2n-1}\right)$ and a
line of fixed points $\mathcal{L}_1:\ \left(\Sigma_*,0\right)$ where
$\Sigma_*\geq 0$ ($\Sigma_*<0$ would imply imaginary shear)
\footnote{$\Sigma_*$ are the coordinates on the $\Sigma$-axis.}. The
point $\Sigma_*=0$ on $\mathcal{L}_1$ represents another isotropic
fixed point that merges with $\mathcal{A}$ when $n=5/4$.

The fixed points may be used to find exact solutions for the Bianchi
I models. We substitute the definitions \rf{DS:var} into
\rf{R:gen_B1} to obtain
\begin{equation}\label{DS:Theta}
\dot{\Th}=\left(\frac{n}{n-1}y_i-\Sigma_i-2\right)\frac{\Th^2}{3},
\end{equation}
where $(\Sigma_i,y_i)$ represents the coordinates of the fixed
points. Given that $n \neq 1$ and $n
y_i-(n-1)\left(\Sigma_i+2\right)\neq 0$, this equation can be
integrated to give
\begin{equation}\label{DS:sol_gen}
a=a_0\left(t-t_0\right)^\alpha,\ \ \ {\rm where} \ \ \
\alpha=\left(2+\Sigma_i-\sfrac{n}{n-1}y_i\right)^{-1}.
\end{equation}

In the case of the fixed point $\mathcal{A}$ we have
\begin{equation}\label{DS:sol_A}
a=a_0\left(t-t_0\right)^{\sfrac{(1-n)(2n-1)}{(n-2)}},
\end{equation}
which is the same as the solution found in \cite{Carloni05}.

For the fixed line $\mathcal{L}_1$ we have
\begin{equation}\label{DS:sol_L1}
a=a_0\left(t-t_0\right)^{\sfrac{1}{2+\Sigma_*}},
\end{equation}
but direct substitution into the cosmological equations reveals that
this solution is only valid for $n>1$. For $n<1$ the fixed points on
$\mathcal{L}_1$ are non physical because the field equations do not
hold there. We have summarised these solution in Table
\ref{Table:vac_sol}.

\begin{table}[tbp] \centering
\caption{The fixed points and eigenvalues for $R^n$-gravity in a LRS
Bianchi I vacuum model.}
\begin{tabular}{lll}
& & \\
\hline  & Fixed points $(\Sigma,y)$ & Eigenvalues \\ \hline
& & \\
Point $\mathcal{A}$ & $\left(0,\sfrac{4n-5}{2n-1}\right)$ &
$\left[\sfrac{2(5-4n)}{n-1},\sfrac{(5-4n)}{n-1}\right]$  \\
& & \\
Line $\mathcal{L}_1$ & $\left(\Sigma_*,0\right)$ &
$\left[0,\sfrac{(4n-5)}{n-1}+\sfrac{(2n-1)}{n-1}\Sigma_*\right]$ \\
& & \\
\hline
\end{tabular}\label{Table:vac_eigen}
\end{table}

\begin{table}[tbp] \centering
\caption{The solutions of the scale factor and shear evolution for
$R^n$-gravity in a LRS Bianchi I vacuum model.}
\begin{tabular}{lll}
& &  \\
\hline  & Scale factor & Shear
\\ \hline
& & \\
Point $\mathcal{A}$ &  $a=a_0\left(t-t_0\right)^{\sfrac{(1-n)(2n-1)}{(n-2)}}$ & $\sigma=0$ \\
& & \\
Line $\mathcal{L}_1$ &
 $a=a_0\left(t-t_0\right)^{\sfrac{1}{2+\Sigma_*}}$,\ \ \ (only valid for $n>1$) & $\sigma=\sigma_0 a^{-(2+\Sigma_*)}$ \\
& & \\
\hline
\end{tabular}\label{Table:vac_sol}
\end{table}

The analysis above would be incomplete without determining the fixed
points at infinity. In order for us to compactify the phase space,
we transform our coordinates $(\Sigma,y)$ to polar coordinates
\begin{equation}
\Sigma=\bar{r}\cos \phi, \ \ \ \ \ y=\bar{r}\sin \phi
\end{equation}
and set $\bar{r}=\frac{r}{1-r}$. Now since $\Sigma \geq 0$, we will
only consider half of the phase space, i.e. $-\pi/2\leq \phi \leq
\pi/2$. In the limit $r \rightarrow 1$ ($\bar{r} \rightarrow
\infty$), equations \rf{DS:eqn2} take on the form
\begin{eqnarray}
r'&&=  \frac{(2n-1)}{4(n-1)}\left[\; \cos \phi -\cos 3\phi -5\sin \phi -\sin 3\phi \; \right], \label{DS:inf_r} \\
\phi ' &&= \frac{(2n-1)\left[\cos \phi -\cos 3\phi +\sin \phi +\sin
3\phi \; \right]}{(n-1)(1-r)}. \label{DS:inf_phi}
\end{eqnarray}

Since \rf{DS:inf_r} does not depend on $r$ we can find the fixed
points by making use of \rf{DS:inf_phi} only. Setting $\phi '=0$ we
obtain four fixed points which are listed in Table 3 with their
corresponding solutions.

\begin{table}[tbp] \centering
\caption{Asymptotic fixed points, $\phi$-coordinates and solutions
for $R^n$-gravity in a LRS Bianchi I vacuum model.}
\begin{tabular}{llll}
& & & \\
\hline  Point & $\phi$ & Scale factor & Shear
\\ \hline
& &  & \\
$\mathcal{A}_\infty$ & $0$
&$\tau-\tau_{\infty}=\frac{1}{\Sigma_c}\ln \left|t-t_0\right|+C$ &
$\sigma=\sigma_0 a^{-(2+\Sigma_c)}$ \\
$\mathcal{B}_\infty$ & $\sfrac{\pi}{2}$ &
$|\tau-\tau_\infty|=\left[C_1 \pm C_0\left|\sfrac{n-1}{2n-1} \right|(t-t_0)\right]^\frac{2n-1}{n-1}$ & $\sigma=0$ \\
$\mathcal{C}_\infty$ & $\sfrac{3\pi}{2}$ &
$|\tau-\tau_\infty|=\left[C_1 \pm C_0\left|\sfrac{n-1}{2n-1}
\right|(t-t_0)\right]^\frac{2n-1}{n-1}$ & $\sigma=0$ \\
$\mathcal{D}_\infty$ & $\sfrac{7\pi}{4}$ &
$|\tau-\tau_\infty|=\left[C_1 \pm C_2(t-t_0)\right]^2$ & $\sigma=\sigma_0$ \\
& & & \\ \hline
\end{tabular}\label{Table:vac_asymp_fix}
\end{table}

The form of the scale factor can be determined from the fixed points
by integrating \rf{DS:inf_r} to find \cite{Clifton05,Holden98}
\begin{equation}
r-1=g(\phi_i)(\tau-\tau_{\infty}),
\end{equation}
where $g(\phi_i)$ represents the right hand side of \rf{DS:inf_r}
and $\tau\to \tau_{\infty}$ as $r \to 1$. The evolution equation
\rf{DS:Theta} is then transformed into polar coordinates
\begin{equation}
\frac{\Th'}{\Th}=\frac{r}{1-r}\left(\frac{n}{n-1}\sin \phi_i-\cos
\phi_i-\frac{2(1-r)}{r}\right),
\end{equation}
which in the limit $r \to 1$ take on the form
\begin{equation}\label{DS:Th_polar}
\frac{\Th'}{\Th} =
\frac{-1}{g(\phi_i)(\tau-\tau_{\infty})}\left(\frac{n}{n-1}\sin
\phi_i-\cos \phi_i\right).
\end{equation}
The equations above were all given in terms of the new time variable
$\tau$ by using the relation $\Th'=3\dot{\Th}/\Th$. Integrating
\rf{DS:Th_polar} yields the solution
\begin{equation}\label{DS:asymp_sol_vac}
|\tau-\tau_{\infty}|=\left[C_1 \pm
C_0\left|h(\phi_i)\right|(t-t_0)\right]^\sfrac{1}{h(\phi_i)},
\end{equation}
where
\begin{equation}
h(\phi_i)=\frac{1}{g(\phi_i)}\left(\frac{n}{n-1}\sin \phi_i-\cos
\phi_i\right)+1.
\end{equation}
Solutions at infinity can now be obtained by directly substituting
the fixed points into \rf{DS:asymp_sol_vac}, so for points
$\mathcal{B}_\infty$ and $\mathcal{C}_\infty$ we find
\begin{equation}\label{DS:sol_inf_BC}
|\tau-\tau_{\infty}|=\left[C_1 \pm
C_0\left|\frac{n-1}{2n-1}\right|(t-t_0)\right]^{\frac{2n-1}{n-1}},
\end{equation}
and for point $\mathcal{D}_\infty$
\begin{equation}\label{DS:inf_sol_D}
|\tau-\tau_{\infty}|=\left[C_1 \pm C_2(t-t_0)\right]^2.
\end{equation}
The solution at point $\mathcal{A}_\infty$ can not be determined
with this method since the limit approaches our fixed line
$\mathcal{L}_1$ so that \rf{DS:asymp_sol_vac} yields an indefinite
solution. However, defining the two new variables: $S=\ln \Sigma$
and $Y=\ln y$, the system \rf{DS:eqn2} can be written as
\begin{eqnarray}\label{DS:eqn_SY}
S' &=& -2\left(\frac{2n-1}{n-1}\right)\; e^Y, \nonumber \\
Y'&=&\left(\frac{2n-1}{n-1}\right)\left(e^S-e^Y\right)+\left(\frac{4n-5}{n-1}\right).
\end{eqnarray}
Point $\mathcal{A}_\infty$ corresponds to $y\to 0$ as $\Sigma\to
\infty$, so in its neighborhood the system \rf{DS:eqn_SY} reduces to
\begin{equation}
S'=0\ \ \ \ {\rm and} \ \ \ \ Y'=\left(\frac{2n-1}{n-1}\right)e^S,
\end{equation}
which has the solution
\begin{equation}
S=S_c= constant\ \ \ \ {\rm and} \ \ \ \
Y=\left(\frac{2n-1}{n-1}\right)e^{S_c}\left(\tau-\tau_\infty\right).
\end{equation}
The form of the scale factor for $\mathcal{A}_\infty$ can then be
found in the same way as the previous points. For $y\to 0$ as
$\Sigma\to \infty$, \rf{DS:Theta} takes the form
$\dot{\Th}=-\Sigma_c\frac{\Th^2}{3}$ where $\Sigma_c=e^{S_c}$. We
find the solution as
\begin{equation}\label{DS:inf_sol_A}
\tau-\tau_{\infty}=\frac{1}{\Sigma_c}\ln \left|t-t_0\right|+C,
\end{equation}
where $C$ is a constant of integration.

\subsection{Stability of the fixed points}

The stability of the fixed point may be determined by linearising
the system of equation \rf{DS:eqn2}. This can be done by perturbing
$\Sigma$ and $y$ around the fixed points $(\Sigma_i,y_i)$ via
$\Sigma=\Sigma_i+\delta \Sigma$ and $y=y_i+\delta y$. The
corresponding eigenvalues of the linearised system are given in
Table \ref{Table:vac_eigen}. The fixed point $\mathcal{A}$ is an
unstable node (repeller) for values of $n$ in the range $1<n<5/4$.
For all other values of $n$ it is a stable node (attractor).

The fixed points on line $\mathcal{L}_1$ all contain at least one
zero eigenvalue and therefore we will have to study the effect of
small perturbations around the line. We find that they have the
following solutions
\begin{equation}\label{DS:per_L1_sol}
\delta \Sigma=\frac{\kappa}{\eta}e^{\eta \tau}, \hspace{10mm} \delta
y=C\; e^{\eta \tau},
\end{equation}
where $C$ is a constant of integration and
\begin{equation}
\eta=\frac{(2n-1)}{(n-1)}\Sigma_*+\frac{(4n-5)}{(n-1)}, \ \ \ \ \ \
\kappa=-\frac{2(2n-1)}{(n-1)}.
\end{equation}
In order for the fixed points on line $\mathcal{L}_1$ to be stable
node, we must have $\eta<0$. The fixed point is an unstable node
when $\eta>0$. Over the interval $0\leq\Sigma_*<\frac{5-4n}{2n-1}$,
we will have stable nodes for $1<n<5/4$ and unstable nodes for
$1/2<n<1$. The remainder of the points $\Sigma_*>\frac{5-4n}{2n-1}$,
will be stable nodes for $1/2<n<1$ and unstable nodes for $1<n<5/4$.
When $n<1/2$ and $n>5/4$, the fixed points are always unstable
nodes. We also note that for $\Sigma_*=1$, $\eta=6$ and is therefore
always an unstable node. The stability of all the fixed points is
given in Table 4.

\begin{table}[tbp] \centering
\caption{Stability of the fixed points for $R^n$-gravity in a LRS
Bianchi I vacuum model. We use the term `attractor' to denote a
stable node and `repeller' refers to an unstable node.}
\begin{tabular}{lcccc}
& & & & \\
\hline  & $n<1/2$& $1/2<n<1$ & $1<n<5/4$ & $n>5/4$
\\ \hline
& & & & \\
Point $\mathcal{A}$ & attractor & attractor & repeller & attractor
\\
& & & &\\\hline
& & & &\\
Line $\mathcal{L}_1$ &  &  &  & \\
$\Sigma_*=1$ & repeller & repeller &
repeller & repeller  \\
$0\leq\Sigma_*<\frac{5-4n}{2n-1}$ & repeller &
repeller & attractor & repeller \\
$\Sigma_*>\frac{5-4n}{2n-1}$ & repeller & attractor &
repeller &  repeller  \\
& & & &\\ \hline
\end{tabular}\label{Table:vac_stab}
\end{table}

A similar analysis may be performed for the fixed points at
infinity. We only need to perturb the angular variable $\phi$ around
the fixed points $\phi_i$ via $\phi=\phi_i+\delta \phi$. The fixed
points will be stable if $r'>0$ and the eigenvalue $\lambda<0$ for
the linearised equation $\delta \phi'= \lambda\; \delta \phi$, in
the limit of $\bar{r}\to \infty$. When both conditions are satisfied
the point is an stable node, if only one is satisfied it is a saddle
and when neither holds it is an unstable node. Substituting the
expression above into \rf{DS:inf_phi} and linearising as before,
yields
\begin{equation}
\hspace{-6mm} \delta \phi' \approx
\frac{(2n-1)}{4(n-1)(1-r)}\left[\; -\sin \phi_i +3\sin 3\phi_i +\cos
\phi_i +3\cos 3\phi_i \; \right]\; \delta \phi .
\end{equation}
The stability of the fixed points are summarised in Table 5. We see
that only point $\mathcal{D}_\infty$ have stable nodes for $n<1/2$
and $n>1$. Points $\mathcal{B}_\infty$ and $\mathcal{C}_\infty$ are
always saddle points and $\mathcal{A}_\infty$ is a saddle when
$1/2<n<1$ but is otherwise an unstable node.

\begin{table}[tbp] \centering
\caption{Stability of the asymptotic fixed points for $R^n$-gravity
in a LRS Bianchi I vacuum model. We use the term `saddle' to denote
a saddle node.}
\begin{tabular}{lccc}
& & & \\
\hline  Point & $n<1/2$ & $1/2<n<1$ & $n>1$
\\ \hline
& & & \\
$\mathcal{A}_\infty$ & repeller &
saddle & repeller \\
$\mathcal{B}_\infty$ & saddle &
saddle & saddle \\
$\mathcal{C}_\infty$ & saddle & saddle & saddle
\\
$\mathcal{D}_\infty$ & attractor &
repeller & attractor \\
& & & \\
\hline
\end{tabular}\label{Table:vac_stab_asymp}
\end{table}

\subsection{Evolution of the shear}

In the previous section we found two isotropic points; the fixed
point $\mathcal{A}$ and one point on the fixed line at $\Sigma_*=0$.
The remaining fixed points all have non-vanishing shear.

The trace free Gauss Codazzi equation \rf{sigdot:Rn_LRS_B1} can in
general (i.e. for all points in the phase space) be represented in
terms of the dynamical variables \rf{DS:var} as
\begin{equation}\label{DS:shear_gen}
\frac{\dot{\sig}}{\sigma}= -\sfrac{1}{3}(2+\Sigma+y)\Th.
\end{equation}
From the equation above it is clear that the shear evolution for all
points in the phase space that lie on the line $y=1-\Sigma$, is the
same as in the case of GR. The shear will dissipate faster than in
GR when $\dot{\sig}/\sig<-\Th$, that is all points that lie in the
region $y<1-\Sigma$. We will call this the {\it fast shear
dissipation} (FSD) regime. When $\dot{\sig}/\sig>-\Th$ and hence for
all points in the region $y>1-\Sigma$, the shear will dissipate
slower than in GR. This will be called the {\it slow shear
dissipation} (SSD) regime.

The fixed points on $\mathcal{L}_1$ for which $\Sigma_*>0$ all have
non-vanishing shear. For these points \rf{DS:shear_gen} has the form
\begin{equation}\label{DS:shear}
\frac{\dot{\sig}}{\sigma}=
-\sfrac{1}{3}(2+\Sigma_*)\Th=-(2+\Sigma_*)\left(\frac{\dot{a}}{a}\right),
\end{equation}
which may be integrated to give
\begin{equation}\label{DS_sol_shear}
\sigma=\sigma_0 a^{-(2+\Sigma_*)}=\sigma_0 a_0^{-(2+\Sigma_*)}
(t-t_0)^{-1},
\end{equation}
where we made use of \rf{DS:sol_L1}. We note that the final solution
of the shear for these fixed points \rf{DS_sol_shear}, does not
depend on the parameter $n$. This is to be expected since both the
coordinates of the points on $\mathcal{L}_1$ and equation
\rf{DS:shear_gen} are independent of $n$.

The only other fixed point with non-vanishing shear is
$\mathcal{D}_\infty$ which corresponds to $y\to -\infty$ as
$\Sigma\to \infty$. In this limit \rf{DS:shear_gen} yields
$\dot{\sigma}/\sigma =0$ which implies that $\sig=\sig_0$ (i.e.
constant shear).

We first consider values of the parameter for which $n<1/2$ (see
Figure 1). If the initial conditions of the universe lie in the
region $y>0$ (negative Ricci scalar), the orbits will always
approach the isotropic fixed point $\mathcal{A}$. The shear will
dissipate slower than in GR for almost all the orbits in this
region, apart from the ones below the dotted line $y=1-\Sigma$,
which make the transition from the SSD region to the FSD region.
When the initial conditions lie in the region $y<0$ (positive Ricci
scalar), the orbits will approach the fixed point
$\mathcal{D}_\infty$ which has constant shear.

The case $1/2<n<1$ is illustrated in Figures 2 and 3. If the initial
conditions are such that they lie in the shaded area ($y<1-\Sigma$
and $y<0$), then the evolution will always be in the FSD regime and
will approach the isotropic solution of the point $\mathcal{A}$.
Instead, the unshaded area $y<0$ and $y>1-\Sigma$, is divided into
two regions; the first one is located below the dash-dotted line and
the second above the dash-dotted line. For initial conditions that
lie in the first region, the orbits make a transition from the SSD
region to the FSD region where they approach the point
$\mathcal{A}$. When the initial conditions lie in the second region,
the orbits will always lie in the SSD region and approach
$\mathcal{L}_1$. When the initial conditions lie in $y<1-\Sigma$ and
$y>0$, the universe will evolve from the FSD regime to the SSD
regime, in which the evolution will approach the stable solutions on
$\mathcal{L}_1$. In the remaining area where $y>1-\Sigma$ and $y>0$,
the shear will always dissipate slower than in GR.

We next consider $1<n<5/4$ which is illustrated in Figure 4. If the
initial conditions of the universe are such that they lie in the
shaded area, then the shear will always dissipate faster than in the
case of GR. When the initial conditions lie in the region
$y>1-\Sigma$ and $y>0$, the universe will evolve from the SSD regime
to the FSD regime and the evolution will approach the stable
solutions on $\mathcal{L}_1$. If the initial conditions lie in the
unshaded area of $y<0$, the orbits will approach the fixed point
$\mathcal{D}_\infty$. For initial conditions that lie in the region
$y<1-\Sigma$, there will be a transition from the FSD region to the
SSD region. For all initial conditions that lie in the region
$y<1-\Sigma$, the orbits always lie in the SSD region. We can see
that for this range of $n$, the fixed point $\mathcal{A}$ acts as a
past attractor. This is an interesting feature since an isotropic
past attractor implies that unlike GR, where the generic
cosmological singularity is anisotropic, we have initial conditions
which corresponds to a FLRW spacetime. This feature was also found
in the braneworld scenarios where it was shown that homogeneous and
anisotropic braneworld models (and some simple inhomogeneous models)
have FLRW past attractors (see e.g.
\cite{Coley02a,Coley02b,Dunsby04,Goheer04}). This means that
although inflation is still required to produce the fluctuations
observed in the cosmic microwave background (CMB), there is no need
for special initial conditions for inflation to begin
\cite{Goode85}. In the range $1<n<5/4$, we can obtain models whose
evolution starts at the isotropic point $\mathcal{A}$ and then
either evolves toward the fixed points ($\Sigma_*,0$) on line
$\mathcal{L}_1$ or towards the point $\mathcal{D}_\infty$. The
orbits that approach $\mathcal{L}_1$ will always lie in the FSD
region. The orbits which approach $\mathcal{D}_\infty$ will make a
transition from the FSD region to the SSD region. An interesting set
of orbits are the ones that approach the fixed points on
$\mathcal{L}_1$ for which $(\sigma/H)_*<<1$. These cosmic histories
represent an universe that is initially isotropic and then develops
shear anisotropies which approach a constant value that can be
chosen to be comparable with the expansion normalised shear observed
today ($(\sigma/H)_*< 10^{-9}$ \cite{Bunn96,Kogut97,Jaffe05})
\footnote{Strictly speaking these orbits do not satisfy the Collins
and Hawking \cite{Collins73} definition for isotropisation, which
require $\sigma/H$ to asymptotically approach zero.}. Inflation is
therefore not required to explain the low degree of anisotropy
observed in the CMB. Furthermore, we still require all other
observational constraints to be satisfied.

Finally, we consider values in the range $n>5/4$ (see Figure 5). For
all initial conditions that lie in the shaded region, the shear will
always dissipate faster than in the case of GR; for $y>0$ the orbits
will approach $\mathcal{A}$ and for $y<0$ approach the point
$\mathcal{D}_\infty$. If initial conditions lie in the region
$y>1-\Sigma$ and $y>0$, the orbits will initially be in the SSD
region, and then approach the isotropic solution $\mathcal{A}$,
which is in the FSD region. For all orbits in the region
$y>1-\Sigma$ and $y<0$, the shear will always dissipate slower than
in the case of GR.

\begin{figure}[tbp]
\begin{center}
\epsfig{file=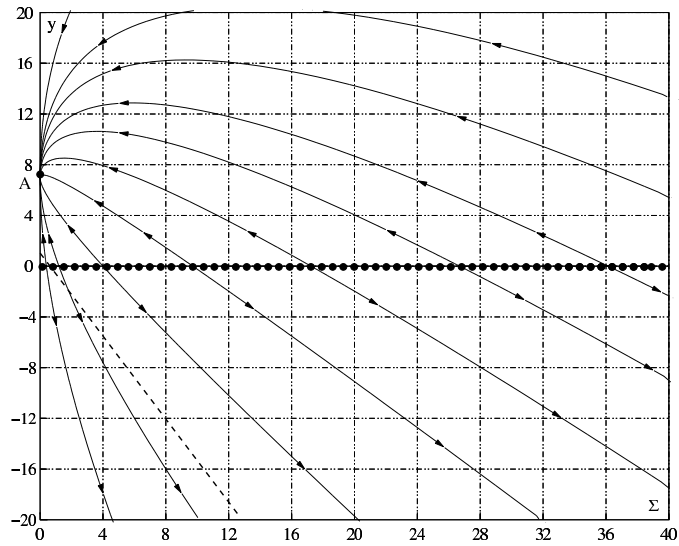, scale=0.6} \caption{Phase space of the
vacuum LRS Bianchi model with $n<1/2$.}
\end{center}\label{fig:n_0.2}
\end{figure}

\begin{figure}[tbp]
\begin{center}
\epsfig{file=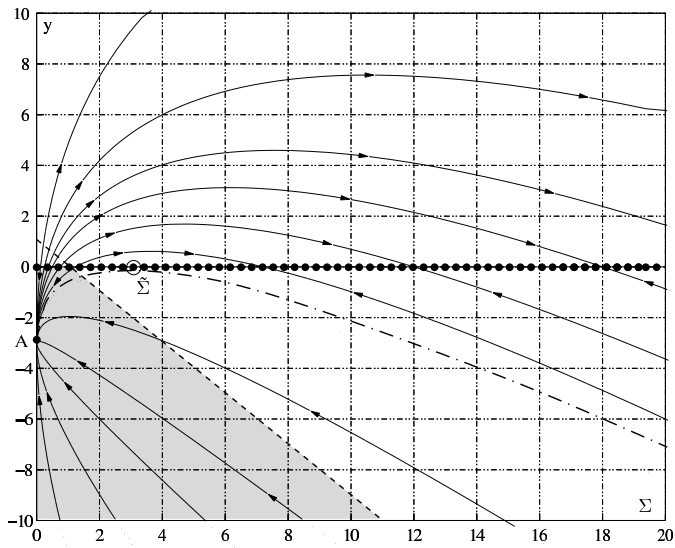, scale=0.6} \caption{Phase space of the
vacuum LRS Bianchi model with $1/2<n<1$ and where
$\tilde{\Sigma}=\frac{5-4n}{2n-1}$. The shaded region represents the
region of initial conditions for which the shear will always evolve
faster than in GR.}
\end{center}\label{fig:n_0.8a}
\end{figure}

\begin{figure}[tbp]
\begin{center}
\epsfig{file=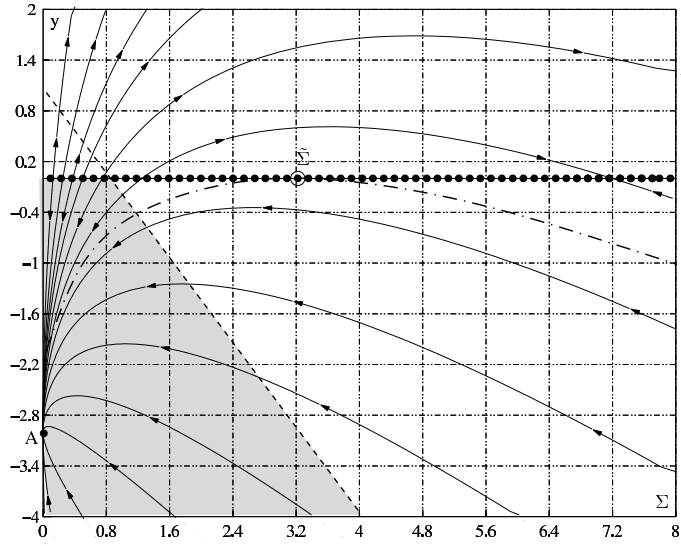, scale=0.6} \caption{Close-up of the
phase space of the vacuum LRS Bianchi model with $1/2<n<1$ around
the line $y=1-\Sigma$.}
\end{center}\label{fig:n_0.8b}
\end{figure}

\begin{figure}[tbp]
\begin{center}
\epsfig{file=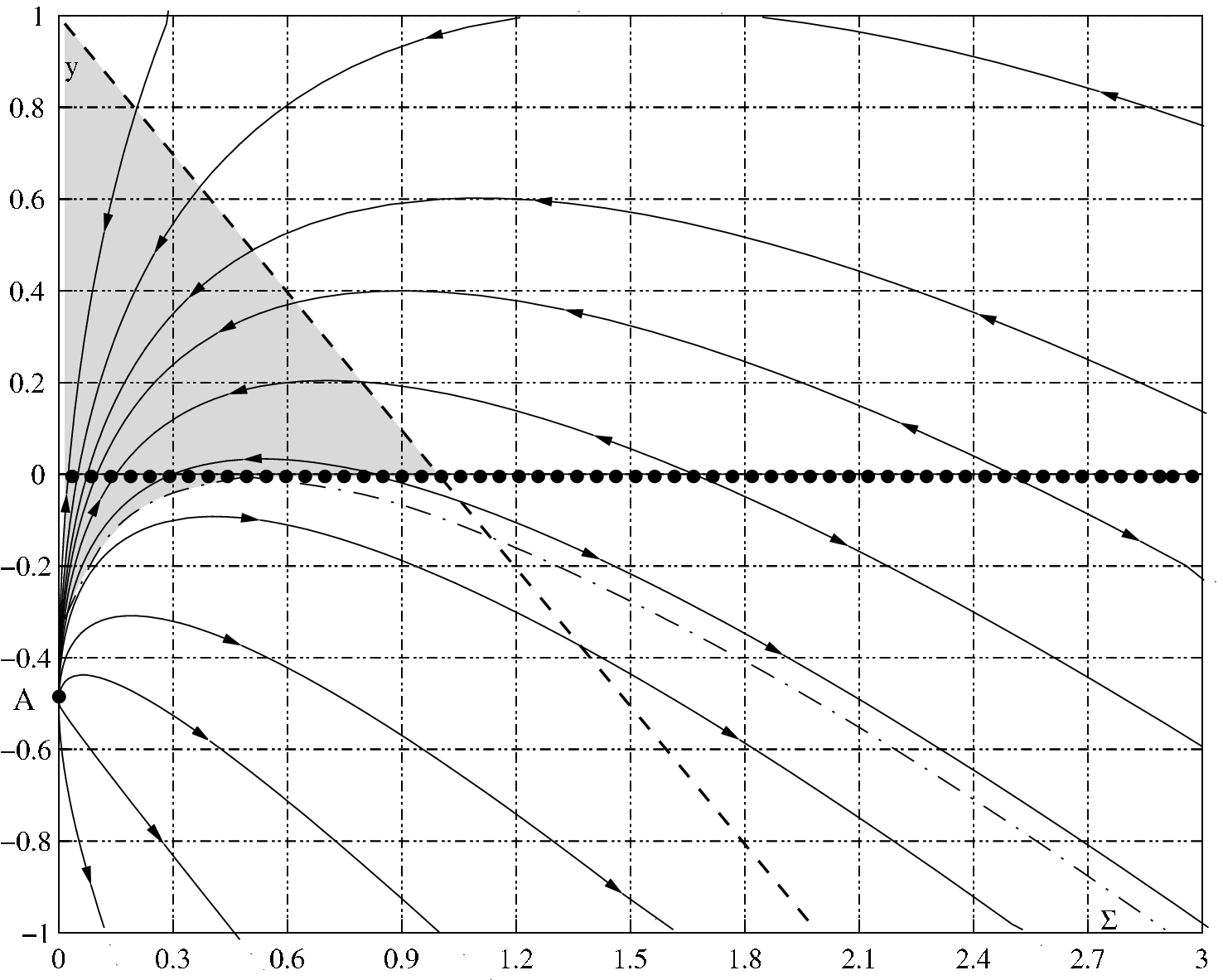, scale=0.6} \caption{Phase space of the
vacuum LRS Bianchi model with $1<n<5/4$.}
\end{center}\label{fig:n_1.1}
\end{figure}

\begin{figure}[tbp]
\begin{center}
\epsfig{file=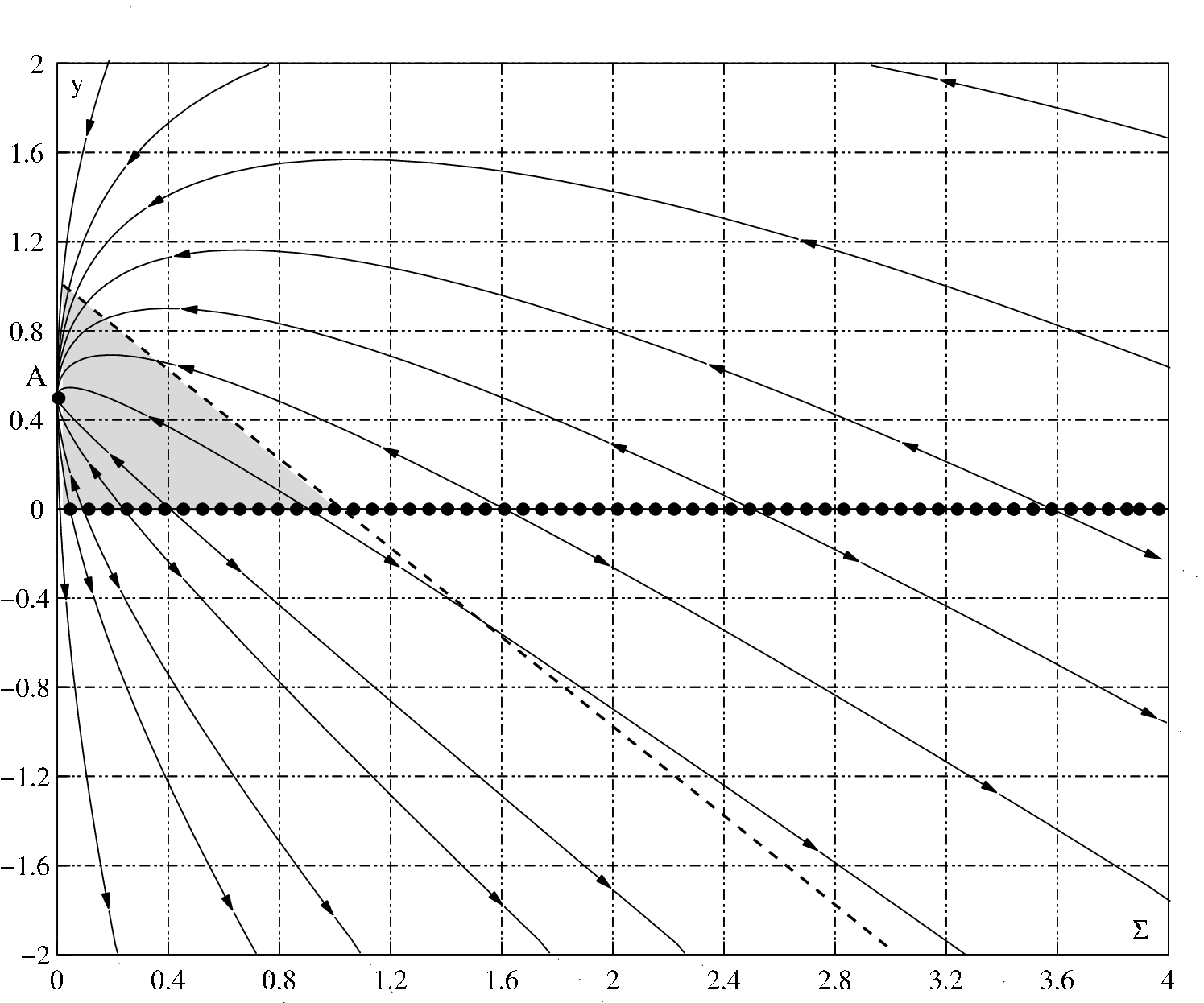, scale=0.6} \caption{Phase space of the
vacuum LRS Bianchi model with $n>5/4$.}
\end{center}\label{fig:n_1.5}
\end{figure}

\section{Dynamics of the matter case}

We will now consider the dynamics of LRS Bianchi I models in the
presence of matter. As noted in \cite{Carloni05}, in HOTG there is a
difference between vacuum and non-vacuum physics in the sense that
not all higher order couplings are consistent in the presence of
standard matter. This can be seen in the evolution equations
\rf{Ray:R^n_B1} and \rf{3R:Rn_B1} where the matter terms are coupled
with a generic power of the curvature. Since the sign of the Ricci
scalar is not fixed, these terms will not be defined for every real
value of $n$. Thus, the inclusion of matter induces a natural
constraint through the field equations on $R^n$- gravity and it is
therefore necessary to express the results in terms of the allowed
set of values of $n$. Following \cite{Carloni05}, we will work as if
$n$ is unconstrained, supposing that the intervals we devise are
meant to represent the subset of allowed values within these
intervals.

Similar to the vacuum case, we can reduce the system \rf{DS:eqn_mat}
to the three variables $\Sigma$, $y$ and $z$ by making use of the
constraint \rf{DS:constraint_mat};
\begin{eqnarray}\label{DS:eqn_mat2}
\hspace{-10 mm} \Sigma'=-2\left[\left(\frac{2n-1}{n-1}\right)\;y+z\right]\Sigma, \nonumber \\
\hspace{-10 mm}y'=\frac{y}{n-1}\left[(2n-1)\Sigma-(2n-1)y+z+(4n-5)\right], \\
\hspace{-10 mm} z' =
z\left[(2-3w)-z+\Sigma-\left(\frac{3n-1}{n-1}\right)\;y\right].
\nonumber
\end{eqnarray}
We note that when $y=0$ then $y'=0$ and when $z=0$, $z'=0$. The two
planes $y=0$ and $z=0$ therefore corresponds to two invariant
submanifolds. When $z=0$, the system \rf{DS:eqn_mat2} reduces to
system \rf{DS:eqn2} and one would be tempted to consider the plane
$z=0$ as the vacuum invariant submanifold of the phase space, which
would not be entirely correct. To illustrate this point, we write
the energy density in terms of our expansion normalised variables
\rf{DS:var}
\begin{equation}\label{DS:mu}
\mu \propto zy^{n-1}\Th^{2n}.
\end{equation}
From this relation it can be seen that when $z=0$ and $y \neq 0$ the
energy density is zero. However when $y=0$ and $z \neq 0$ the
behaviour of $\mu$ does depend on the value of $n$. In this case the
energy density is zero when $n>1$ but is divergent when $n<1$. When
both $y$ and $z$ are equal to zero and $n<1$, one can only determine
the behaviour of $\mu$ by direct substitution into the cosmological
equations.

\subsection{Fixed points and solutions}

Setting $\Sigma'=0$, $y'=0$ and $z'=0$ we obtain three isotropic
fixed points $\mathcal{A}$, $\mathcal{B}$, $\mathcal{C}$ and a line
of fixed points $\mathcal{L}_1:\ \left(\Sigma_*,\ 0,\ 0\right)$,
where $\Sigma_*\geq 0$ (see Table \ref{mat_fixed}). When
$\Sigma_*=0$ we have another isotropic fixed point which merges with
$\mathcal{A}$ when $n=5/4$ and with $\mathcal{B}$ when $w=2/3$. This
point will merge with $\mathcal{C}$ when $n=5/4$ and $w=2/3$.

We again substitute the definitions \rf{DS:var} into \rf{R:gen_B1}
to obtain
\begin{equation}\label{DS:Theta_mat}
\dot{\Th}=\left(\frac{n}{n-1}y_i-\Sigma_i-2\right)\frac{\Th^2}{3}.
\end{equation}
Under the condition that $n \neq 1$ and the terms inside the
brackets are not equal to zero, this equation may be integrated to
give the following solution
\begin{equation}\label{DS:sol_gen_mat}
a=a_0\left(t-t_0\right)^\alpha,\ \ \ {\rm where} \ \ \
\alpha=\left(2+\Sigma_i-\sfrac{n}{n-1}y_i\right)^{-1}.
\end{equation}
The point $\mathcal{A}$ and line $\mathcal{L}_1$ will all have the
same solutions as in the vacuum case, since either $y=0$ or $z=0$
for these points.

The behaviour of the scale factor for point $\mathcal{B}$ is
\begin{equation}\label{DS:sol_B_mat}
a=a_0\left(t-t_0\right)^{1/2},
\end{equation}
and since $y_2=0$ and $z_2 \neq0$, the energy density is zero (only
valid for $n>1$). When $n<1$, point $\mathcal{B}$ and the points on
$\mathcal{L}_1$ are non physical since the energy density is
divergent.

For point $\mathcal{C}$, the scale factor behaves as
\begin{equation}\label{DS:sol_C_mat}
a=a_0\left(t-t_0\right)^{\sfrac{2n}{3(1+w)}},
\end{equation}
while the energy density is
\begin{equation}\label{DS:mu_C}
\mu=\mu_0 t^{-2n},
\end{equation}
where
\begin{eqnarray*} \mu_0 &=& (-1)^n 3^{-n} 2^{2n-1} n^n
(1+w)^{-2n}(4n-3(1+w))^{n-1} \\
&& \times \left[2n^2(4+3w)-n(13+9w)+3(1+w)\right].
\end{eqnarray*}
This point thus represents a power-law regime which in the case of
$n>0$, yields an expanding solution with the energy density
decreasing in time. In the case of $n<0$ we obtain a contracting
solution with $\mu$ increasing in time. In order for $\mathcal{C}$
to be a physical point, we require $\mu>0$ and therefore $\mu_0>0$
(see \cite{Carloni05} for detailed analysis). Note than when
$n>\sfrac{3}{2}(1+w)$, this solution corresponds to accelerated
expansion.

\begin{table}[tbp] \centering
\caption{Fixed points and solutions of the scale factor and matter
density for LRS Bianchi I with matter.}
\begin{tabular}{llll}
& & &\\
\hline Point  & Fixed points $(\Sigma,y,z)$ & Scale factor & Matter
density
\\ \hline
& & & \\
$\mathcal{A}$ &
$\left(0,\;\sfrac{4n-5}{2n-1},\;0\right)$ &  $a=a_0\left(t-t_0\right)^{\sfrac{(1-n)(2n-1)}{(n-2)}}$ & $\mu=0$\\
$\mathcal{B}$ &
$\left(0,\; 0,\; 2-3w\right)$ &  $a=a_0\left(t-t_0\right)^{\frac{1}{2}}$ & $\mu=0$ \ \ ($n>1$)\\
$\mathcal{C}$ & $\left(0,\
\frac{(n-1)[4n-3(1+w)]}{2n^2},\right.$ &  $a=a_0\left(t-t_0\right)^{\sfrac{2n}{3(1+w)}}$ & $\mu=\mu_0 t^{-2n}$\\
& $\left.\frac{n(13+9w)-2n^2(4+3w)-3(1+w)}{2n^2}\right)$& & \\
& & & \\
Line $\mathcal{L}_1$ &  $\left(\Sigma_*,0,0\right)$ &
$a=a_0\left(t-t_0\right)^{\sfrac{1}{2+\Sigma_*}}$ & $\mu=0$\ \ ($n>1$)\\
& & & \\
\hline
\end{tabular}\label{mat_fixed}
\end{table}

We next study the behaviour of the system \rf{DS:eqn_mat2} at
infinity. The compactification of the phase space can be achieved by
transforming to spherical coordinates
\begin{equation}
\Sigma=\bar{r}\sin \theta \cos \phi, \ \ \ \ y=\bar{r}\sin \theta
\sin \phi, \ \ \ \ z=\bar{r} \cos \theta,
\end{equation}
and setting $\bar{r}=\frac{r}{1-r}$, where $0\leq \bar{r} \leq
\infty$, $0\leq \theta \leq \pi$ and since we are again only
considering half of the phase space, $-\pi/2\leq \phi \leq \pi/2$ .
In the limit $r \rightarrow 1$ ($\bar{r} \rightarrow \infty$),
equations \rf{DS:eqn_mat2} take on the form
\begin{eqnarray}
\hspace{-15 mm} r' = \frac{1}{4(n-1)}\left[\;2\cos
\theta\{3-2n+(2n-1)
\cos 2\phi\}+2\cos^3 \theta \{(2n-1) \cos 2\phi-1\}\right. \nonumber \\
\hspace{5 mm} +2\sin \theta \;\cos^2 \theta \{(2n-1) \cos 2\phi-1\}
\{ \cos \phi
+\sin \phi\} \nonumber \\
\hspace{5 mm} +(2n-1)\sin \theta \left.\{ \cos \phi -\cos 3\phi -5\sin \phi -\sin 3\phi\}  \; \right], \label{DS:inf_r_mat} \\
\hspace{-15 mm} \theta' = \frac{\sin 2\theta \{1-(2n-1) \cos
2\phi\}\left[\cos \theta +\sin \theta\{ \cos \phi +\sin \phi\} \;
\right]}{4(n-1)(1-r)}
  \label{DS:inf_th_mat} \\
\hspace{-15 mm} \phi ' = \frac{(2n-1)\sin 2\phi \left[\cos \theta
+\sin \theta\{ \cos \phi +\sin \phi\} \; \right]}{2(n-1)(1-r)}.
\label{DS:inf_phi_mat}
\end{eqnarray}

\begin{table}[tbp] \centering
\caption{Eigenvalues and shear solutions for LRS Bianchi I with
matter.}
\begin{tabular}{lll}
& & \\
\hline Point &  Eigenvalues & Shear
\\ \hline
&  & \\
$\mathcal{A}$ &
$\left[\sfrac{2(5-4n)}{n-1},\; \sfrac{5-4n}{n-1},\; \sfrac{n(13+9w)-2n^2(4+3w)-3(1+w)}{1 - 3n + 2n^2} \right]$ & $\sigma=0$ \\
$\mathcal{B}$&
$\left[-2+ 3 w,\; -4+6w,\; \sfrac{4n-3(1+w)}{n-1} \right]$ & $\sigma=0$ \\
$\mathcal{C}$&
$\left[\sfrac{3((2n-1)w-1)}{n},\; \sfrac{P_1(n,w)-\sqrt{P_2(n,w)}}{4n(n-1)},\; \sfrac{P_1(n,w)+\sqrt{P_2(n,w)}}{4n(n-1)} \right]$ & $\sigma=0$ \\
& $P_1(n,w)= 3(1+w)+3n\left( (2n-3)w-1\right) $ & \\
& $P_2(n,w)= (n-1)\left[4n^3 (8+3w)^2-4n^2(152+3w(55+18w))\right. $ & \\
& \hspace{15mm} $\left.+3n(1+w)(139+87w)-81(1+w)^2 \right]$ & \\
& & \\
Line $\mathcal{L}_1$ &  $\left[0,\; \sfrac{(4n-5)}{n-1}+\sfrac{(2n-1)}{n-1}\Sigma_*,\; 2-3w+\Sigma_*\right]$ & $\sigma=\sigma_0 a^{-(2+\Sigma_*)}$\\
& & \\ \hline
\end{tabular}\label{mat_eigen}
\end{table}

Now since \rf{DS:inf_r_mat} does not depend on $r$ we can find the
fixed points by just making use of \rf{DS:inf_th_mat} and
\rf{DS:inf_phi_mat}. Setting $\theta'=0$ and $\phi '=0$ we obtain
the fixed points which are listed in Table \ref{mat_fixed_asymp}
with their corresponding solutions.

\begin{table}[tbp] \centering
\caption{Coordinates and solutions of the asymptotic fixed points
for LRS Bianchi with matter. For the fixed line
$\tilde{\theta}=\arctan[-1/(\cos \phi_i+\sin \phi_i)]$ and for the
double point $\tilde{\phi}=(1/2)\arccos[1/(2n-1)]$. }
\begin{tabular}{llll}
& & & \\
\hline  Point & $(\theta,\phi)$  & Scale factor & Shear
\\ \hline
& & & \\
 $\mathcal{A}_\infty$ &$(0,0)$   & $|\tau-\tau_{\infty}|=\left[C_1 \pm C_0(t-t_0)\right]$ & $\sigma=0$ \\
$\mathcal{B}_\infty$ & $(\pi,0)$ & $|\tau-\tau_{\infty}|=\left[C_1
\pm C_0(t-t_0)\right]$ & $\sigma=0$
 \\
$\mathcal{C}_\infty$ & $(\sfrac{\pi}{2},0)$  &
$\tau-\tau_{\infty}=\frac{1}{\Sigma_c}\ln \left|t-t_0\right|+C$ &
$\sigma=\sigma_0 a^{-(2+\Sigma_c)}$ \\
$\mathcal{D}_\infty$ & $(\sfrac{\pi}{2},\sfrac{\pi}{2})$ &
$|\tau-\tau_\infty|=\left[C_1 \pm C_0\left|\sfrac{n-1}{2n-1} \right|(t-t_0)\right]^\frac{2n-1}{n-1}$  & $\sigma=0$ \\
$\mathcal{E}_\infty$ & $(\sfrac{\pi}{2},\sfrac{3\pi}{2})$ &
$|\tau-\tau_\infty|=\left[C_1 \pm C_0\left|\sfrac{n-1}{2n-1}
\right|(t-t_0)\right]^\frac{2n-1}{n-1}$ &
$\sigma=0$ \\
$\mathcal{F}_\infty$ & $(\tilde{\theta},\tilde{\phi})$ &
$|\tau-\tau_\infty|=\left[C_1 \pm C_2(t-t_0)\right]^2$  & $\sigma=\sigma_0$\\
& & & \\
Line & & & \\\hline
$\mathcal{L}_\infty$& $(\tilde{\theta},\phi)$ & $|\tau-\tau_\infty|=\left[C_1 \pm C_2(t-t_0)\right]^2$ & $\sigma=\sigma_0$\\
& & & \\
\hline
\end{tabular}\label{mat_fixed_asymp}
\end{table}

The form of the scale factor can  be determined from the fixed
points as in the vacuum case. We integrate \rf{DS:inf_r_mat} to find
\begin{equation}
r-1=G(\theta_i,\phi_i)(\tau-\tau_{\infty}),
\end{equation}
where $G(\theta_i,\phi_i)$ represents the right hand side of
\rf{DS:inf_r_mat} and $\tau\to\tau_{\infty}$ as $r\to1$. We
transform the evolution equation \rf{DS:Theta_mat} to polar
coordinates and write them in terms of the time parameter $\tau$:
\begin{equation}
\frac{\Th'}{\Th}=\frac{r\;\sin \theta_i}{1-r}\left(\frac{n}{n-1}\sin
\phi_i-\cos \phi_i-\frac{2(1-r)}{r\;\sin \theta_i}\right),
\end{equation}
which in the limit $r \to 1$ takes the form
\begin{equation}\label{DS:Th_polar_mat}
\frac{\Th'}{\Th} = \frac{-\sin
\theta_i}{G(\theta_i,\phi_i)(\tau-\tau_{\infty})}\left(\frac{n}{n-1}\sin
\phi_i-\cos \phi_i\right).
\end{equation}
Integrating the equation above then yields
\begin{equation}\label{DS:asymp_sol_mat}
|\tau-\tau_{\infty}|=\left[C_1 \pm
C_0\left|H(\theta_i,\phi_i)\right|(t-t_0)\right]^\sfrac{1}{H(\theta_i,\phi_i)},
\end{equation}
where
\begin{equation}
H(\theta_i,\phi_i)=\frac{\sin
\theta_i}{G(\theta_i,\phi_i)}\left(\frac{n}{n-1}\sin \phi_i-\cos
\phi_i\right)+1.
\end{equation}
The solutions is then obtained by directly substituting the
coordinates of the fixed points into \rf{DS:asymp_sol_mat}. For
points $\mathcal{A}_\infty$ and $\mathcal{B}_\infty$ we have the
solutions
\begin{equation}\label{DS:inf_sol_mat_AB}
|\tau-\tau_{\infty}|=\left[C_1 \pm C_0(t-t_0)\right],
\end{equation}
and for point $\mathcal{D}_\infty$ and $\mathcal{E}_\infty$ we have
\begin{equation}\label{DS:inf_sol_mat_DE}
|\tau-\tau_{\infty}|=\left[C_1 \pm
C_0\left|\frac{n-1}{2n-1}\right|(t-t_0)\right]^{\frac{2n-1}{n-1}}.
\end{equation}
In addition to the ordinary fixed points, we have a line of fixed
points $\mathcal{L}_\infty$ for which
$\tilde{\theta}=\arctan[-1/(\cos \phi_i+\sin \phi_i)]$, and a double
fixed point at $\tilde{\phi}=(1/2)\arccos[1/(2n-1)]$. Equation
\rf{DS:asymp_sol_mat} gives the solutions for the fixed points on
the line:
\begin{equation}\label{DS:inf_sol_mat_L}
|\tau-\tau_{\infty}|=\left[C_1 \pm C_2(t-t_0)\right]^2.
\end{equation}

As in the vacuum case, $\mathcal{C}_\infty$ can not be determined
with this method since the limit approaches the fixed line
$\mathcal{L}_1$. In order to analyse this point, we can define three
new variables: $S=\ln \Sigma$, $Y=\ln y$ and $Z=\ln z$. The system
\rf{DS:eqn_mat2} can then be written as
\begin{eqnarray}\label{DS:eqn_mat_C}
S' &=& -2\left(\frac{2n-1}{n-1}\right)\; e^Y-2e^Z, \nonumber \\
Y'&=&\left(\frac{2n-1}{n-1}\right)\left(e^S-e^Y\right)+\frac{1}{n-1}e^Z+\left(\frac{4n-5}{n-1}\right),
\\
Z' &=& (2-3w) -e^Z+e^S-\left(\frac{3n-1}{n-1}\right)e^Y. \nonumber
\end{eqnarray}
The point $\mathcal{C}_\infty$ corresponds to  $y\to 0$ and $z\to 0$
as $\Sigma\to \infty$ so that the system \rf{DS:eqn_mat_C} reduce to
\begin{equation}
S'=0, \ \ \ \ \ Y'=\left(\frac{2n-1}{n-1}\right)e^S,\ \ \ \ \
Z'=e^S,
\end{equation}
which has the following solution
\begin{equation}
S=S_c, \ \ \
Y=\left(\frac{2n-1}{n-1}\right)e^{S_c}\left(\tau-\tau_\infty\right),\
\ \ {\rm and}\ \ \ Z=e^{S_c}\left(\tau-\tau_\infty\right),
\end{equation}
where $S_c$ is a constant. The form of the scale factor for
$\mathcal{C}_\infty$ can then be found in the same way as the
previous points. For $y\to 0$ and $z\to 0$ as $\Sigma\to \infty$,
\rf{DS:Theta_mat} takes the form
$\dot{\Th}=-\Sigma_c\frac{\Th^2}{3}$ where $\Sigma_c=e^{S_c}$. We
find the solution as
\begin{equation}\label{DS:inf_sol_mat_C}
\tau-\tau_{\infty}=\frac{1}{\Sigma_c}\ln \left|t-t_0\right|+C,
\end{equation}
where $C$ is a constant of integration.

\subsection{Stability of the fixed points}

We next check the stability of the fixed point by linearising the
system of equation \rf{DS:eqn_mat2}. The eigenvalues of the
linearised system are given in Table~7.

For the stability analysis, we consider three cases: dust $w=0$,
radiation $w=1/3$ and stiff matter $w=1$. The stability of the fixed
points $\mathcal{A}$ and $\mathcal{B}$ are summarised in Tables
\ref{mat_stab_A} and \ref{mat_stab_B} respectively. Their behaviour
is similar to the flat ($k=0$) points ($\mathcal{C}$ and
$\mathcal{F}$) considered in \cite{Carloni05}. The stability
analysis for the fixed point $\mathcal{C}$ cannot be performed in an
exact way. The eigenvalues are complex in the following ranges: When
$w=0$, for $0.31\lesssim n \lesssim 0.71$ and $1\lesssim n \lesssim
1.32$; when $w=1/3$, for $0.35\lesssim n \lesssim 1.28$; and when
$w=1$, for $0.37\lesssim n \lesssim 1$ and $1.224\lesssim n \lesssim
1.47$. The results are given in Table \ref{mat_stab_C}.

\begin{table}[tbp] \centering
\caption{Stability of the fixed point $\mathcal{A}$ for LRS Bianchi
I with matter. The parameters are
$N_{\pm}=\sfrac{1}{16}(13\pm\sqrt{73})$,
$P_{\pm}=\sfrac{1}{5}(4\pm\sqrt{6})$ and
$Q_{\pm}=\sfrac{1}{14}(11\pm\sqrt{37})$.}
\begin{tabular}{lllll}
& & & &\\
\hline  & {\small $n< N_{-}$} & {\small $N_{-}< n < P_{-}$} &
{\small $P_{-}< n < Q_{-}$} & {\small $Q_{-}< n < 1/2$}
\\ \hline
{\small $w=0$} & {\small attractor} &
{\small saddle} & {\small saddle} & {\small saddle}\\
{\small $w=1/3$} & {\small attractor} &
{\small attractor} & {\small saddle} & {\small saddle}\\
{\small $w=1$} & {\small attractor} & {\small attractor} & {\small
attractor} & {\small saddle}
\\
& & & & \\
& {\small $1/2<n<1$} & {\small $1<n<Q_{+}$} & {\small $Q_{+}< n <
5/4$ }& {\small $5/4 < n < P_{+}$}
\\ \hline
{\small $w=0$} & {\small attractor} &
{\small repeller} & {\small repeller} & {\small saddle}\\
{\small $w=1/3$} & {\small attractor} &
{\small repeller} & {\small repeller} &  {\small saddle} \\
{\small $w=1$} & {\small attractor} & {\small repeller} & {\small saddle} & {\small attractor} \\
& & & & \\
& {\small $P_{+}< n< N_{+}$} & {\small $n> N_{+}$} & & \\
\hline {\small $w=0$} & {\small saddle} & {\small attractor} &
 & \\
{\small $w=1/3$} & {\small attractor} & {\small attractor} &
&  \\
{\small $w=1$} & {\small attractor} & {\small attractor} &  &
\\
\hline
\end{tabular}\label{mat_stab_A}
\end{table}

\begin{table}[tbp] \centering
\caption{Stability of the fixed point $\mathcal{B}$ for LRS Bianchi
I with matter.}
\begin{tabular}{ccccc}
& & & &\\
\hline  & $n<3/4$ & $3/4<n<1$ & $1<n<3/2$ & $n>3/2$
\\ \hline
& & & & \\
$w=0$ & saddle &
attractor & saddle & saddle\\
$w=1/3$ & saddle &
saddle & saddle & saddle\\
$w=1$ & repeller & repeller & saddle & repeller
\\
& & & & \\
\hline
\end{tabular}\label{mat_stab_B}
\end{table}

\begin{table}[tbp] \centering
\caption{Stability of the fixed point $\mathcal{C}$ for LRS Bianchi
I with matter. We use the term `spiral' to denote a pure attractive
spiral and an `anti-spiral' refers to a pure repulsive spiral.}
\begin{tabular}{llllll}
& & & &\\
\hline  & {\small $0<n\lesssim 0.31$} & {\small $0.31\lesssim n
\lesssim 0.35$} & {\small $0.35\lesssim n \lesssim 0.37$} & {\small
$0.37\lesssim n \lesssim 0.71$} & {\small $0.71\lesssim n \lesssim
0.75$}
\\ \hline
{\small $w=0$} & {\small attractor} &
{\small spiral} & {\small spiral} & {\small spiral} & {\small attractor}\\
{\small  $w=1/3$} & {\small attractor} &
{\small attractor} & {\small spiral} & {\small spiral} & {\small spiral} \\
{\small $w=1$} & {\small saddle} & {\small saddle} & {\small saddle}
& {\small spiral} & {\small spiral}
\\
& & & & &\\
& {\small $0.75\lesssim n\lesssim 1$} & {\small  $1\lesssim n
\lesssim 1.220$} & {\small  $1.220 < n \lesssim 1.224$} & {\small
$1.224\lesssim n\lesssim 1.28$} & {\small $1.28\lesssim n\lesssim
1.29$}
\\ \hline
{\small $w=0$} & {\small saddle} &
{\small spiral} & {\small spiral} & {\small spiral} & {\small spiral}\\
{\small $w=1/3$} & {\small spiral} &
{\small spiral} & {\small spiral} &  {\small spiral} & {\small attractor}\\
{\small $w=1$} & {\small spiral} & {\small saddle} & {\small repeller} & {\small anti-spiral} & {\small anti-spiral}\\
& & & & &\\
& {\small $1.29\lesssim n\lesssim 1.32$}
 & {\small $1.32\lesssim n\lesssim 1.34$} & {\small $1.34\lesssim n\lesssim 1.47$} & {\small $1.47\lesssim n\lesssim 1.5$} &
{\small $n\gtrsim 1.5$}
\\ \hline
{\small $w=0$} & {\small spiral} & {\small attractor} & {\small
saddle}
 & {\small saddle} & {\small saddle} \\
{\small $w=1/3$} & {\small saddle} & {\small saddle} & {\small
saddle}
& {\small saddle} & {\small saddle} \\
{\small $w=1$} & {\small anti-spiral} & {\small anti-spiral} &
{\small anti-spiral} & {\small repeller} & {\small saddle}
\\
\hline
\end{tabular}\label{mat_stab_C}
\end{table}

As in the vacuum case we find that the fixed points on the line
$\mathcal{L}_1$ have zero eigenvalues. We therefore study the
perturbations around the fixed line which lead to the following
solutions
\begin{equation}\label{DS:per_mat_L1_sol}
\delta \Sigma=- \sfrac{2(2n-1)}{(n-1)}\sfrac{C_0}{\eta}e^{\eta
\tau}-2\Sigma_*\;\sfrac{C_1}{\kappa} e^{\kappa \tau}, \ \ \ \ \delta
y=\sfrac{C_0}{\eta} e^{\eta \tau}, \ \ \ \ \delta
z=\sfrac{C_1}{\kappa} e^{\kappa \tau},
\end{equation}
where $C_0$ and $C_1$ are constants of integration and
\begin{equation}
\eta=\frac{(2n-1)}{(n-1)}\Sigma_*+\frac{(4n-5)}{(n-1)}\;, \ \ \ \ \
\ \ \ \kappa=2-3w+\Sigma_*.
\end{equation}
In order for the fixed points on line $\mathcal{L}_1$ to be stable
nodes, we must have $\eta<0$ and $\kappa<0$. When $\eta>0$ and
$\kappa<0$ or $\eta<0$ and $\kappa>0$ we have a saddle and when
$\eta>0$ and $\kappa>0$ it is an unstable node. The results have
been summarised in Table \ref{mat_stab_L}.

\begin{table}[tbp] \centering
\caption{Stability of the fixed line $\mathcal{L}_1$ for LRS Bianchi
I with matter. The coordinates for all these points are
$(\Sigma_*,0,0)$ and therefore we only state the range of values on
the fixed line in terms of their $\Sigma$-coordinates.}
\begin{tabular}{cccc}
& & & \\
\hline {\small $w=0,\;1/3$} & {\small Attractors} & {\small Saddles} & {\small Repellers}  \\
\hline
& & & \\
{\small $n<1/2$} & {\small none} & {\small none} & {\small
$\Sigma_*>0$}
\\
{\small $1/2<n<1$} & {\small none} & {\small
$\Sigma_*>\frac{5-4n}{2n-1}$} & {\small
$0<\Sigma_*<\frac{5-4n}{2n-1}$}
\\
{\small $1<n<2$} & {\small none}& {\small
$0<\Sigma_*<\frac{5-4n}{2n-1}$} & {\small
$\Sigma_*>\frac{5-4n}{2n-1}$}
\\
{\small $n>2$} & {\small none} & {\small none} & {\small
$\Sigma_*>0$}
\\
& & & \\\hline {\small $w=1$} & & & \\\hline
& & & \\
{\small $n<1/2$ }& {\small none} & {\small $0<\Sigma_*<1$} & {\small
$\Sigma_*>1$}
\\
{\small $1/2<n<1$} & {\small none} &
$\left\{\begin{tabular}{l}{\small $\Sigma_*>\frac{5-4n}{2n-1}$} \\
{\small $0<\Sigma_*<1$}
\end{tabular}\right.$ & {\small $1<\Sigma_*<\frac{5-4n}{2n-1}$}
\\
{\small $1<n<5/4$} & {\small $0<\Sigma_*<\frac{5-4n}{2n-1}$} &
{\small $\frac{5-4n}{2n-1}<\Sigma_*<1$} & {\small $\Sigma_*>1$}
\\
{\small $n>5/4$} & {\small none} & {\small $0<\Sigma_*<1$} & {\small
$\Sigma_*>1$}
\\
\hline
\end{tabular}\label{mat_stab_L}
\end{table}

A similar analysis can be performed for the fixed points at
infinity. We can check the stability of the fixed point by
linearising the system of equation
\rf{DS:inf_r_mat}-\rf{DS:inf_phi_mat}. The eigenvalues of the
linearised system are given in Table~\ref{mat_eigen_asymp}. The
stability of the fixed points $\mathcal{A}_\infty$,
$\mathcal{B}_\infty$, $\mathcal{C}_\infty$, $\mathcal{D}_\infty$ and
$\mathcal{E}_\infty$ can then be found straightforwardly as in the
vacuum case (see Table \ref{mat_stab_asymp}). Points
$\mathcal{A}_\infty$ and $\mathcal{B}_\infty$ are always saddle
points. The point $\mathcal{C}_\infty$ lie on the fixed line
$\mathcal{L}_1$ and is an unstable node for $n<1/2$ and $n>1$ and a
saddle for $1/2<n<1$. Point $\mathcal{D}_\infty$ is an unstable node
for $n<0$ and $n>1$, a saddle for $0<n<1/2$ and a stable node for
$1/2<n<1$. Point $\mathcal{E}_\infty$ is a stable node for $n<0$ and
$n>1$, a saddle for $0<n<1/2$ and an unstable node for $1/2<n<1$.

\begin{table}[tbp] \centering
\caption{Eigenvalues and value of $r'$ of the ordinary asymptotic
fixed points for LRS Bianchi I with matter. }
\begin{tabular}{lllc}
& & & \\
\hline  Point & $(\theta,\phi)$ & Eigenvalues & $r'$
\\ \hline
& & & \\
 $\mathcal{A}_\infty$ &$(0,0)$ & $\left[-1,\sfrac{2n-1}{n-1} \right]$ & -1  \\
$\mathcal{B}_\infty$ & $(\pi,0)$ & $\left[1,-\sfrac{2n-1}{n-1}
\right]$ & 1  \\
$\mathcal{C}_\infty$ & $(\sfrac{\pi}{2},0)$ &
$\left[1,\sfrac{2n-1}{n-1} \right]$ & 0 \\
$\mathcal{D}_\infty$ & $(\sfrac{\pi}{2},\sfrac{\pi}{2})$ &
$\left[-\sfrac{n}{n-1},\sfrac{2n-1}{n-1} \right]$ &
$-\sfrac{2n-1}{n-1}$  \\
$\mathcal{E}_\infty$ & $(\sfrac{\pi}{2},\sfrac{3\pi}{2})$ &
$\left[-\sfrac{2n-1}{n-1},\sfrac{n}{n-1} \right]$ &
$\sfrac{2n-1}{n-1}$ \\
& & &  \\
\hline
\end{tabular}\label{mat_eigen_asymp}
\end{table}

\begin{table}[tbp] \centering
\caption{Stability of the ordinary asymptotic fixed points for LRS
Bianchi I with matter. Results are independent of $w$.}
\begin{tabular}{lcccc}
& & & & \\
\hline  Point & $n<0$ & $0<n<1/2$ & $1/2<n<1$ & $n>1$
\\ \hline
& & & & \\
$\mathcal{A}_\infty$ & saddle &
saddle & saddle & saddle \\
$\mathcal{B}_\infty$ & saddle &
saddle & saddle & saddle\\
$\mathcal{C}_\infty$ & repeller & repeller & saddle
& repeller \\
$\mathcal{D}_\infty$ & attractor & saddle & saddle & attractor\\
$\mathcal{E}_\infty$ & repeller &
saddle & attractor & repeller  \\
& & & & \\
\hline
\end{tabular}\label{mat_stab_asymp}
\end{table}

The eigenvalues of the fixed line  $\mathcal{L}_\infty$ are given
by
\begin{equation}
[\lambda_1,\lambda_2]= \left[\frac{P_1(n,\phi)-\sqrt{
P_2(n,\phi)}}{4(n-1)(2+\sin
2\phi)^{3/2}},\frac{P_1(n,\phi)+\sqrt{P_2(n,\phi)}}{4(n-1)(2+\sin
2\phi)^{3/2}}\right],
\end{equation}
where
\begin{eqnarray*}
P_1(n,\phi)&=&(3n-4)\cos \phi+n \cos 3\phi-(3n+1)\sin \phi+(n-1)\sin
3
\phi, \\
P_2(n,\phi)&=&2(2+\sin 2\phi)^{3/2}\left[(2n^2-2n+1)(1+\sin 2
\phi)\right. \\
&&\left.+(2n-1)(-\cos 2\phi-2 \sin 4\phi +(2n-1)\sin 6 \phi)\right].
\end{eqnarray*}
The stability can then be found in a similar fashion as the ordinary
asymptotic fixed points. The eigenvalues in this case is dependent
on two variables, $n$ and $\phi$, which makes it difficult to
express the results in a table. We have therefore summarised these
results in a diagram (see Figure~6) \footnote{This diagram was found
by plotting $\tau=\lambda_1+\lambda_2$, $\Delta=\lambda_1 \lambda_2$
and $\tau^2-4\Delta$ and using the definitions for stability to
classify the regions \cite{Strogatz}.}. The stability of any fixed
point on the line $\mathcal{L}_\infty$ for a given value of $n$ can
be read from this diagram. For example the black dot in Figure~6
represents the fixed point at $\phi=0.4$ for a model with $n=1.4$.
It lies within a region that classify it as an attractor.

\begin{figure}[tbp]
\begin{center}
\epsfig{file=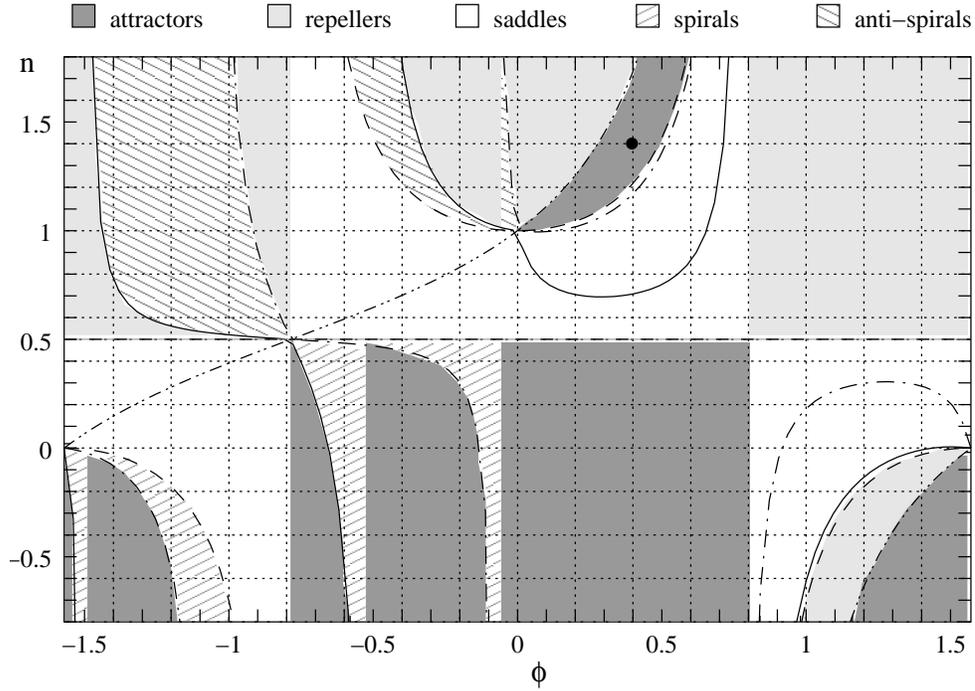} \caption{Stability of the line of
fixed points $\mathcal{L}_\infty$ for LRS Bianchi I with matter.
Results are independent of $w$. The black dot represents the fixed
point at $\phi=0.4$ for a model with $n=1.4$.}
\end{center}\label{fig:stab_Linf}
\end{figure}

\subsection{Evolution of the shear}

The trace free Gauss Codazzi equation \rf{sigdot:Rn_LRS_B1} can in
general (i.e. for all points in the phase space) be represented in
terms of the dynamical variables \rf{DS:var} as
\begin{equation}\label{DS:shear_gen_mat}
\frac{\dot{\sig}}{\sigma}= -\sfrac{1}{3}(2+\Sigma+y+z)\Th.
\end{equation}
The shear evolves at the same rate as in GR when
$\dot{\sig}/\sig=-\Th$, which holds for all values on the plane
$1=\Sigma+y+z$. The shear will dissipate faster than in GR when
$\dot{\sig}/\sig<-\Th$, that is all points that lie in the region
$1>\Sigma+y+z$. We may again call this the {\it fast shear
dissipation} (FSD) regime. The shear will dissipate slower than in
GR when $\dot{\sig}/\sig>-\Th$ and hence all points in the region
$1<\Sigma+y+z$. This will be called the {\it slow shear dissipation}
(SSD) regime. Analysing the three dimensional phase space for this
system is more difficult than the two dimensional spaces considered
in the vacuum case since it is harder to visualise.

All the finite fixed points ($\mathcal{A}$, $\mathcal{B}$ and
$\mathcal{C}$) together with the point $\Sigma_*=0$ on
$\mathcal{L}_1$, lie on the plane $\Sigma=0$ and are therefore
isotropic.

The evolution of the shear for the anisotropic fixed points on
$\mathcal{L}_1$ can be obtained as in the vacuum case. For these
points \rf{DS:shear_gen_mat} take the form
\begin{equation}\label{DS:shear_m}
\frac{\dot{\sig}}{\sigma}=
-\sfrac{1}{3}(2+\Sigma_*)\Th=-(2+\Sigma_*)\left(\frac{\dot{a}}{a}\right).
\end{equation}
which may be integrated to give
\begin{equation}\label{DS_sol_shear_m}
\sigma=\sigma_0 a^{-(2+\Sigma_*)}=\sigma_0
a_0^{-(2+\Sigma_*)}(t-t_0)^{-1},
\end{equation}
where we made use of \rf{DS:sol_gen}. This is the same solutions
that where obtained in the vacuum case. This is to be expected since
all these fixed points lie on the line $\mathcal{L}_1$ for which
$z=0$.

The point $\mathcal{C}_\infty$ and the points on the line
$\mathcal{L}_\infty$ (including $\mathcal{F}_\infty$) are the only
asymptotic fixed points with non-vanishing shear. Point
$\mathcal{C}_\infty$ lies on the line $\mathcal{L}_1$ and therefore
has the solution
\begin{equation}\label{DS_sol_shear_C_inf}
\sigma=\sigma_0 a^{-(2+\Sigma_c)}.
\end{equation}
The behaviour of the shear for the fixed points on
$\mathcal{L}_\infty$ can be found like \rf{DS:Th_polar_mat}, i.e by
transforming \rf{DS:shear_gen_mat} into polar coordinates and taking
the limit $r\to 1$. The resulting equation then yields the solution
$\sigma=\sigma_0={\rm constant}$.

\section{Discussion and Conclusions}

We have derived the evolution equations of the shear for Bianchi I
cosmologies with  $f(R)$-gravity. This general expression, allows us
to consider the shear evolution for any function of the scalar
curvature. However, because the shear depend non-linearly on the
Ricci scalar one can not determine how the dissipation of the shear
anisotropy compares with the case in GR even if one chooses a
specific form for $f(R)$ (such as $R+R^2$ or $R^n$). One  way of
dealing with this problem is to make certain assumptions (for
example the form of the evolution of the scale factor $a$
\cite{Maartens94}) to obtain a solution. A more general approach is
to make use of the dynamical systems approach to study HOTG in these
cosmologies since it provides both exact solutions and the global
behaviour of the system.

Our main aim in this paper was to see how the shear behaves in LRS
Bianchi I cosmologies with $R^n$- gravity and whether these models
isotropises at early and late times. To achieve this goal we used
the theory of dynamical systems to analyse the system of equations
governing the evolution of this model with and without matter.

The phase space for these models have a number of interesting
features, in particular it contains one isotropic fixed point and a
line of fixed points with non-vanishing shear. The isotropic fixed
point is an attractor (stable node) for values of the parameter $n$
in the ranges $n<1/2$, $1/2<n<1$ and $n>5/4$. In the range $1<n<5/4$
this point is a repeller (unstable node) and therefore may be seen
as a past attractor. An isotropic past attractor implies that
inflation can start without requiring special initials conditions.
However, since we have attractors for $(\sigma/H)_*<<1$ on
$\mathcal{L}_1$, we may not need inflation since the shear
anisotropy approaches a constant value which may be chosen as the
expansion normalised shear observed today ($(\sigma/H)_*< 10^{-9}$
\cite{Bunn96,Kogut97,Jaffe05}), provided that other observational
constraints such as nucleosynthesis are satisfied.

We also found that the line $y=1-\Sigma$ separates the phase space
into two part. For all points on this line, the shear dissipates at
the same rate as in GR. In the region above the line the shear
dissipates faster (FSD) than GR and in the region below the line,
slower (SSD) than in GR. From Figures $1-5$ we can see that there
are a number of orbits which cross the dotted line. These are
systems which initially lie in the FSD region and then make a
transition to the SSD region and {\it vice versa}. An interesting
feature of the vacuum case is that when the evolution of the
universe reaches the stable solutions on $\mathcal{L}_1$, the shear
will evolve according to $\sig \propto t^{-1}$ irrespective of the
value of $n$. For values that lie in the range $1/2<n<1$ and $y>0$,
one may have orbits that initially lie in the FSD region (see
Figures 1 and 2) and then make a transition to the SSD region at
late times. The opposite will happen for values that lie in the
range $1<n<5/4$ and $y>0$. Initially the orbits lie in the SSD
region (see Figure 3) and then make a transition to the FSD region.

We observe the same kind of behaviour in the matter case where the
phase space is however 3-dimensional, but is similarly divided into
two regions, by the plane $1=\Sigma+y+z$. The space above the plane
is the SSD region and below the FSD region. Similar argument to the
vacuum case can be used here to investigate the orbits. When matter
is included we do however only have stable fixed points on
$\mathcal{L}_1$ for values of $n$ in the range $1<n<5/4$.

In conclusion we have shown that $R^n$- gravity modifies the
dynamics of the shear in LRS Bianchi I cosmologies by altering the
rate at which the shear dissipates. There are cases in which the
shear always dissipate slower or faster than in GR, and there are
ones which make transitions from first evolving faster and later
slower (and {\it vice versa}) than in GR.

\ack

This research was supported by the National Research Foundation
(South Africa) and the Italian {\it Ministero Degli Affari Esteri-DG
per la Promozione e Cooperazione Culturale} under the joint Italy/
South Africa Science and Technology agreement. A special thanks to
the group at the University of Naples, Federico II, for their
hospitality during the early stages of this work. We thank the
referees for their useful comments.

\section*{References}


\begin{thebibliography}{99}

\bibitem{Carroll04} Carroll S M, Duvvuri V, Trodden M and Turner M S
2004 \prd {\bf 70} 043528

\bibitem{Nojiri03} Nojiri S and Odintsov S D 2003 \prd {\bf 68}
123512

\bibitem{Olmo05} Olmo G J 2005 \prd {\bf 72}
083505

\bibitem{Capozzi02} Capozziello S 2002 {\it Int. Journ. Mod. Phys.} D {\bf 11}
483

\bibitem{Capozzi03}Capozziello S, Carloni S and Troisi A 2003 {\it
Recent Res. Devel. Astronomy \& Astrophysics} {\bf 1} 625 ({\it
Preprint} astro-ph/0303041)

\bibitem{Dynamical} Wainwright J and Ellis G F R (ed) 1997 {\it
Dynamical systems in cosmology} (Cambridge: Cambridge University
Press) (see also references therein)

\bibitem{Carloni05} Carloni S, Dunsby P K S, Capozziello S and Troisi
A 2005 \cqg {\bf 22} 4839

\bibitem{Clifton05} Clifton T and Barrow J D 2005 \prd {\bf 72} 103005

\bibitem{Maartens94} Maartens R and Taylor D R 1994 \grg \textbf{26} 599


\bibitem{carge73}  Ellis G F R 1973 \textit{Carg\`{e}se Lectures in Physics,}
\textit{Vol} \textbf{6} (ed) E Scatzman  (New York: Gordon and
Breach)

\bibitem{Cargese}  Ellis G F R   and van Elst H 1999 \textit{Cosmological
Models (Carg\`{e}se Lectures 1998), Theoretical and Observational
Cosmology} (ed) M. Lachi\`{e}ze-Rey (Kluwer, Dordrecht) 1-116 ({\it
Preprint} gr-qc/9812046)

\bibitem{Kasner} Kasner E 1925 {\it Trans. Am. Math. Soc.} {\bf 27}
101

\bibitem{Barrow06} Barrow J D and Clifton T 2006 \cqg {\bf
23} L1

\bibitem{Clifton06a} Clifton T and Barrow J D 2006 \cqg {\bf
23} 2951

\bibitem{Ellis67} Ellis G F R 1967 \jmp {\bf 8} 1171

\bibitem{Stewart68} Stewart J M and Ellis G F R 1968 \jmp {\bf 9}
1072

\bibitem{vElst96}  van Elst H and Ellis G F R 1996 \cqg {\bf
13} 1099

\bibitem{Coley02a} Coley A A 2002 \cqg {\bf 19} L45

\bibitem{Coley02b} Coley A A 2002 \prd {\bf 66} 023512

\bibitem{Dunsby04} Dunsby P K S, Goheer N, Bruni M and Coley A 2004 \prd {\bf
69} 101303

\bibitem{Goheer04} Goheer N, Dunsby P K S, Bruni M and Coley A 2004 \prd {\bf
70} 123517

\bibitem{Rippl96} Rippl S, van Elst H, Tavakol R and Taylor D 1996 \grg
\textbf{28} 193

\bibitem{Berkin90}  Berkin A L 1990 \prd {\bf 42} 1016

\bibitem{Holden98} Holden D J and Wands D 1998 \cqg {\bf 15} 3271

\bibitem{Goode85} Goode S W and Wainwright J 1985 \cqg {\bf 2} 99

\bibitem{Bunn96} Bunn E F, Ferreira P G and Silk J 1996 \prl {\bf
77} 2883

\bibitem{Kogut97} Kogut A, Hinshaw G and Banday A J 1997 \prd {\bf
55} 1901

\bibitem{Jaffe05} Jaffe T R, Banday A J, Eriksen H K, G\'{o}rski K M
a nd Hansen F K 2005 \apj {\bf 629} L1

\bibitem{Collins73} Collins C B and Hawking S W 1973 \apj {\bf 180}
317

\bibitem{Strogatz} Strogatz S H 1994 {\it Nonlinear Dynamics and
Chaos} (Cambridge, Massachusetts: Perseus Books)
\end{thebibliography}
\end{document}